\RequirePackage{fix-cm}
\documentclass[natbib,smallextended]{svjour3}       % onecolumn (second format)
\smartqed  % flush right qed marks, e.g. at end of proof
\usepackage{graphicx}
\usepackage{amsmath}
\usepackage{amssymb}
\usepackage{ulem} % AV
\usepackage{color} % AV
 \journalname{Space Science Reviews}
\newcommand{\lsim}{\lesssim}
\newcommand{\gsim}{\gtrsim}
\newcommand{\levy}{{L\'evy}}
\newcommand{\Ppeak}{p_\mathrm{peak}}
\newcommand{\Dpen}{D_\mathrm{pen}}

\newcommand{\Pxx}{P_\mathrm{xx}}
\newcommand{\Fnum}{F_\mathrm{n}}
\newcommand{\FnumZ}{F_\mathrm{n}^0}
\newcommand{\Fion}{f_\mathrm{ion}}
\newcommand{\Reff}{R_\mathrm{eff}}

\newcommand{\alfL}{\alpha_L}
\newcommand{\alfH}{\alpha_H}

\newcommand{\GRBs}{$\gamma$-ray bursts}

\newcommand{\GBeta}{\gamma(x)\beta(x)}
\newcommand{\GBetaZ}{\gamma_0\beta_0}

\newcommand{\pFEB}{p_\mathrm{FEB}}

\newcommand{\delMax}{\delta\theta_\mathrm{max}}

\newcommand{\delAng}{\delta\theta}

\newcommand{\PAS}{pitch-angle scattering}

\newcommand{\gamZ}{\gamma_{\rm sh}}%{\gamma_0}

\newcommand{\transrel}{trans-rel\-a\-tiv\-is\-tic}
\newcommand{\Transrel}{Trans-rel\-a\-tiv\-is\-tic}
\newcommand{\ultrarel}{ul\-tra-rel\-a\-tiv\-is\-tic}

\newcommand{\Lor}{Lorentz}

\newcommand{\Lmfp}{\lambda_\mathrm{mfp}}

\newcommand{\SSC}{synchrotron-self-Comptonization}
\newcommand{\syn}{synchrotron}

\newcommand{\pion}{pion-decay}
\newcommand{\IC}{inverse Compton}
\newcommand{\ICd}{inverse-Compton}
\newcommand{\brems}{bremsstrahlung} % DCE

\newcommand{\HeT}{He$^{2+}$}

\newcommand{\Valf}{v_{a}}
\newcommand{\xx}[1]{\!\times\!10^{#1}}

\newcommand{\Lfeb}{L_\mathrm{FEB}}

\newcommand{\kmps}{km s$^{-1}$}

\newcommand{\SA}{semi-analytic}
\newcommand{\SAn}{Semi-analytic}

\newcommand{\mfp}{mean free path}

\newcommand{\PIC}{Particle-in-cell}

\newcommand{\Bls}{B_\mathrm{ls}}

\newcommand{\pD}{p_d}

\newcommand{\rg}{r_g}

\newcommand{\rgZ}{r_{g0}}

\newcommand{\gyrotime}{\tau_g}
\newcommand{\deltime}{\delta t}
\newcommand{\etamfp}{\eta_\mathrm{mfp}}

\newcommand{\Vscat}{v_\mathrm{scat}}

\newcommand{\PhiPz}{\Phi_{P0}}

\newcommand{\Fen}{F_\mathrm{en}}
\newcommand{\FenZ}{F_\mathrm{en}^0}
\newcommand{\Fpx}{F_\mathrm{px}}
\newcommand{\nonrel}{non-relativistic}
\newcommand{\FpxZ}{\PhiPz}
\newcommand{\rel}{relativistic}

\newcommand{\pmax}{p_\mathrm{max}}

\newcommand{\mc}{Monte Carlo}

\newcommand{\DSA}{diffusive shock acceleration}
\newcommand{\Facc}{Fermi acceleration}
\newcommand{\FoFSA}{first-order Fermi shock acceleration}

\newcommand{\MFA}{magnetic field amplification}
\newcommand{\alf}{Alfv\'en}

\newcommand{\NL}{nonlinear}
\newcommand{\SCly}{self-consistently}
\newcommand{\SC}{self-consistent}
\newcommand{\SCc}{Self-consistent}

\newcount\listnorom
\newcommand\listromanDE{\global\advance \listnorom by 1
{\lowercase\expandafter{(\romannumeral\listnorom)}\ }}
\newcommand\newlistroman{\listnorom=0}

\newcount\listnumber
\newcommand\listDE{\global\advance \listnumber by 1
{\lowercase\expandafter{(\number\listnumber)}\ }}
\newcommand\newlistDE{\listnumber=0}
\newcount\Inum
\newcount\IInum
\newcount\IIInum
\newcount\IVnum
\def\I{\global\multiply\IInum by 0 \global\multiply\IIInum by 0
            \global\multiply\IVnum by 0 \global\advance \Inum by 1
            {\the\Inum. }}
\def\II{\global\multiply\IIInum by 0\global\multiply\IVnum by 0
       \global\advance \IInum by 1 {\the\Inum.\the\IInum. }}
\def\III{\global\multiply\IVnum by 0\global\advance \IIInum by 1
            {\the\Inum.\the\IInum.\the\IIInum. }}
\def\IV{\global\advance \IVnum by 1
            {\the\IVnum. }}
\begin{document}

\title{Towards understanding the physics of collisionless relativistic shocks
  %\thanks{Grants or other notes
%about the article that should go on the front page should be
%placed here. General acknowledgments should be placed at the end of the article.}
}
\subtitle{Relativistic collisionless shocks}

%\titlerunning{Short form of title}        % if too long for running head

\author{Guy Pelletier          \and Andrei Bykov \and Don Ellison \and
  Martin Lemoine \and}

%\authorrunning{Short form of author list} % if too long for running head

\institute{G. Pelletier \at UJF-Grenoble 1  CNRS-INSU, Institut de Plan\'etologie et
  d'Astrophysique de Grenoble (IPAG) UMR 5274,
  F-38041 Grenoble, France\\  \email{guy.pelletier@obs.ujf-grenoble.fr} % \\
%             \emph{Present address:} of F. Author  %  if needed
  \and
  A. Bykov \at Ioffe Institute for Physics and Technology, 194021 St-Petersburg, Russia\\
  \at International Space Science Institute, CH-3012 Bern, Switzerland\\
  \at St-Petersburg State Polytechnical University, 195251 St-Petersburg, Russia
   \\  \email{ambykov@yahoo.com}
 \and
  D. C. Ellison \at Physics Department, North Carolina State University, Box8202, Raleigh, NC27694, USA
  \\  \email{don\_ellison@ncsu.edu}
  \and
  M. Lemoine \at Institut d'Astrophysique de Paris, UMR-7095
           CNRS - UPMC, 98 bis boulevard Arago, F-75014 Paris, France
  \\  \email{lemoine@iap.fr}
}

\date{Received: date / Accepted: date}
% The correct dates will be entered by the editor

\maketitle

\begin{abstract}
Relativistic astrophysical collisionless shocks represent outstanding
dissipation agents of the huge power of relativistic outflows produced
by accreting black holes, core collapsed supernovae and other objects
into multi-messenger radiation (cosmic rays, neutrinos,
electromagnetic radiation). This article provides a theoretical
discussion of the fundamental physical ingredients of these extreme
phenomena. In the context of weakly magnetized shocks, in particular,
it is shown how the filamentation type instabilities, which develop in
the precursor of pair dominated or electron-ion shocks, provide the
seeds for the scattering of high energy particles as well as the agent
which preheats and slows down the incoming precursor plasma. This
analytical discussion is completed with a mesoscopic, non-linear model
of particle acceleration in relativistic shocks based on
\mc\ techniques. This \mc\ model uses a semi-phenomenological
description of particle scattering which allows it to calculate the
back-reaction of accelerated particles on the shock structure on
length and momentum scales which are currently beyond the range of
microscopic particle-in-cell (PIC) simulations. 
% \PACS{PACS code1 \and PACS code2 \and more}
% \subclass{MSC code1 \and MSC code2 \and more}
\end{abstract}

\section{Introduction}\label{sec:int}
Relativistic collisionless shocks are extreme phenomena of plasma
physics which form at the interface of the relativistic outflows of
powerful astrophysical sources (e.g. gamma-ray bursts, pulsar wind
nebulae and active galactic nuclei) and their environment. These shock
fronts are not only outstanding dissipation agents, which convert into
relativistic heat and disorder a well ordered relativistic flow, they
also appear to produce quasi-power law spectra of accelerated
particles up to very high energies through a relativistic variant of
the Fermi process. These particles either escape the acceleration site
and thus become cosmic rays, possibly very high energy cosmic rays,
and/or they interact with ambient backgrounds to produce high energy
photons and possibly neutrinos over a wide range of energies. In this
regard, the physics of relativistic shocks is a key element in the
arena of astroparticle physics and high energy astrophysics.

How such shocks form, how particle acceleration takes place, how
magnetic turbulence is excited and to what level, remain fundamental
open questions. Reviews on the physics, the numerical simulation and
the phenomenology of relativistic collisionless shocks exist in the
literature, which describe the progress achieved so far, see in
particular \citet{BykovEtal2012,SKL2015,2016RPPh...79d6901M}. The
present review takes a different stance and tries to describe
analytically the physics of these phenomena to some level of detail.
Due to the limited space, not all details could be presented nor could
all regimes be investigated. The emphasis has been deliberately placed
on unmagnetized relativistic shocks, which are representative, for
example, of external shocks in gamma-ray bursts; other types of shocks
are nonetheless briefly addressed to provide a more general picture.

The paper is laid out as follows; in Sec.~\ref{sec:basics}, we provide
essential notions on relativistic collisionless shock waves,
discussing in particular the shock jump conditions and the particle
kinematics; Sec.~\ref{sec:theory} desribes how a relativistic
collisionless shock forms, depending on the ambient magnetization and
shock velocity; Sec.~\ref{sec:prec} discusses the physics of the
precursor, in which the superthermal particle population mixes with
the background plasma, and where the latter is pre-heated and slowed
down; Sec.~\ref{sec:phenom} presents some phenomenological
consequences and observational tests of the theory.  Finally, we
complete this analytical model with a description of \mc\ numerical
simulation techniques in Sec.~\ref{MC1}, which provides a
complementary point of view on the physics of relativistic
shocks. While Monte Carlo simulations must use as input some
microphysics describing particle kinematics, which can be derived from
the analytical theory presented earlier, they also provide a
mesoscopic non-linear picture which is useful to extrapolate
microphysical results to astrophysical scales of interest. Our main
results are summarized in Sec.~\ref{sec:sum}.
Some details of a more technical nature are presented in
Apps.~\ref{sec:appA}, \ref{sec:appB} and \ref{sec:appC}.

\section{The basics of Relativistic Shocks}\label{sec:basics}
Just as any other shock front, a relativistic shock in a plasma
corresponds to a transition between two plasma states at a ``front"
which moves at velocity close to the velocity of light relative to
some unshocked medium (the background plasma). Such a shock transition
requires that the shock front moves at a velocity $v_{\rm sh}$ which
is faster than the velocity at which causal disturbances can be
transmitted in the background unshocked plasma. In the hydrodynamical
limit, this velocity is $c_{\rm s}$, the sound velocity; if the
background plasma is magnetized, however, various modes can propagate
in the plasma, at different velocities, hence several types of shock
waves can be envisaged; see \citet{1999JPhG...25R.163K} for a detailed
discussion. This review will be interested in fast shock waves, which
move faster than the fast magnetosonic velocity $\left(c_{\rm
  s}^2+v_{\rm A}^2\right)^{1/2}$, $v_{\rm A}$ corresponding to the
Alfv\'en velocity in the background plasma. For simplicity, it will
also be assumed that the shock front is planar and that it propagates
into the background plasma along its normal.

Upon crossing the shock front, the ordered kinetic energy is converted
into thermal energy mostly. A description of the physics of shock
fronts is conveniently made in the so-called front frame that
separates two homogeneous plasmas, namely the \emph{upstream plasma},
the unshocked background plasma that inflows into the shock front, and
the \emph{downstream plasma}, which flows away from the shock at a
subsonic velocity as a consequence of intense heating in the shock
transition.  In a collisional plasma, this transition is usually quite
sharp, its size being measured in units of the mean free path of
particles. In collisionless, non-relativistic plasmas, with an oblique
magnetic field (i.e. meaning that the field lines have a finite angle
with respect to the shock normal), the size of the transition is
measured in term of the gyration radius of ions. In parallel shocks
(i.e. with a magnetic field parallel to the shock normal), the
transition is generally much more extended, because strong
disturbances are generated and the absence of collisional dissipation
allows them to travel over larger distances.
In this case, the upstream and downstream plasmas close to the shock
front need not be uniform.  Nevertheless, on some scale, provided the
downstream flow has been isotropized, the so-called Rankine-Hugoniot
jump relations, which express the balance of the fluxes of matter,
momentum and energy, are fulfilled.

In the next section, we recall these jump conditions and discuss the
kinematics of particles around this shock front, which are peculiar to
relativistic shock waves.

\subsection{Shock jump conditions}
\begin{figure}
\begin{center}
\includegraphics[width=1.0\columnwidth]{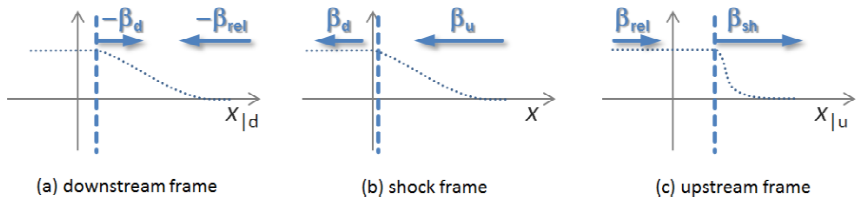}
\caption{Sketch of the shock transition in various frames, along with
  the flow velocities. The dotted line represents a sketch of the
  density distribution of superthermal particles; the extension of
  this distribution in the upstream delimits the shock precursor. In
  the shock rest frame (b), the downstream moves at velocity
  $\beta_{\rm d}$ relative to the shock front, while the upstream is
  incoming at velocity $\beta_{\rm u}$. In the upstream rest frame
  (c), the downstream plasma moves at velocity $\beta_{\rm rel}$,
  while the shock front moves at $\beta_{\rm sh}\,=\,-\beta_{\rm
    u}$. Note that the extension of the precursor is reduced by the
  Lorentz contraction in this rest frame, relative to the shock rest
  frame. Finally, in the downstream rest frame (a), the shock front
  moves away at velocity $-\beta_{\rm d}$, while the usptream is
  moving at $-\beta_{\rm rel}$.}
\label{fig:frames}
\end{center}
\end{figure}

In the case of a relativistic shock, it is convenient to derive these
jump relations in the shock front frame by using the energy-momentum
tensor $T_{\mu \nu}$ \citep[e.g.][]{BM1976,DoubleEtal2004}. A word of
caution regarding notations: except otherwise noted, and except for
proper quantities such as density $n$, energy density $e$ and pressure
$p$ or enthalpy $h$, quantities will be expressed in the shock front
rest frame; if expressed in the upstream or downstream rest frames,
these quantities will be indexed with $_{\rm u}$ or $_{\rm d}$
respectively. Figure~\ref{fig:frames} illustrates the various frames
used in this text and the corresponding flow velocities.

Assuming a plane front and a normal coordinate $x$, the relations
express that the momentum flux $T_x^x$ and the energy flux $T_x^t$
keep the same values upstream and downstream.  The momentum flux
conservation reads:
\begin{equation}
\label{eq:jc0}
(e_{\rm d}+p_{\rm d}) u_{\rm d}^2 + p_{\rm d} = n_{\rm u}m c^2 u_{\rm u}^2 \ ,
\end{equation}
and the energy flux conservation reads:
\begin{equation}
\label{eq:jc1}
(e_{\rm d}+p_{\rm d}) \Gamma_{\rm d} u_{\rm d} =n_{\rm u}mc^2 \Gamma_{\rm u} u_{\rm u}
\end{equation}
$u_{\rm u}$ (resp.  $u_{\rm d}$) denote the $x-$component of the
upstream (resp. downstream) plasma $4-$velocity in the shock frame,
with corresponding Lorentz factor $\Gamma_{\rm u}$ (resp. $\Gamma_{\rm
  d}$). To make contact with our notations above, the shock
velocity relative to the unshocked background plasma is of course
$\beta_{\rm sh}\,=\,-\beta_{\rm u}$, with associated Lorentz factor
$\gamma_{\rm sh}\,=\,\left(1-\beta_{\rm sh}^2\right)^{-1/2}$.

The above equations assume a cold upstream plasma. We introduce an
equation of state $p_{\rm d}= w_{\rm d} e_{\rm d}$ for the downstream
plasma; since the downstream plasma is relativistically hot, $w_{\rm
  d} = 1/3$. Note that this does not require that the distribution
function reflect thermal equilibrium, since this ratio holds for any
isotropic distribution for which most of the particles are
ultra-relativistic; in particular the distribution function can be
composed of a thermal part with a high energy isotropized tail. One
can generalize the above equation of state to that of a plasma
containing a magnetic component set in the transverse directions to
the shock normal (i.e. a magnetic field lying in the shock front
plane), carrying a fraction $\sigma_{\rm d}$ of the energy density of
the downstream plasma. Then,
\begin{equation}
\label{eq:jc2}
w_{\rm d} = \frac{1}{3}\frac{1+ 3\sigma_{\rm d}}{1+ \sigma_{\rm d}} \ .
\end{equation}
Now, the downstream velocity $\beta_{\rm d}$ is obtained by solving the
following algebraic equation derived from the two conservation
equations:
\begin{equation}
\label{eq:jc3}
\beta_{\rm d}^2 - (1+w_{\rm d})\beta_{\rm u} \beta_{\rm d} + w_{\rm d} = 0 \ .
\end{equation}
For $\beta_{\rm u}^2 \,\simeq\, 1$, corresponding to the
ultra-relativistic limit, the two roots are $1$ and $w_{\rm d}$.  The
downstream pressure of a strong relativistic shock is then
\begin{equation}
\label{eq:jc4}
p_{\rm d}=(1-w_{\rm d}) \gamma_{\rm sh}^2 n_{\rm u}m c^2 \ .
\end{equation}

From the mass flux conservation law, one derives easily that the
\emph{apparent} density in the shock front frame is amplified by a
factor $w_{\rm d}^{-1}$, as measured in the front rest-frame; however
the Lorentz transform from the background plasma rest-frame to the
shock front frame amplifies the \emph{proper} density by a factor
$\gamma_{\rm sh}$, while the Lorentz transform from the shock frame to
the downstream plasma rest-frame reduces the proper density by a
factor $\gamma_{\rm d}\,\simeq\,3/(2\sqrt{2})$ in the hydrodynamic
limit $\sigma_{\rm d}\,\ll\,1$. Thus the proper density eventually
suffers an amplification by a compression ratio corresponding to the
jump of the 4-velocity vector $r \equiv u_{\rm u}/u_{\rm d} =
2\sqrt{2} \gamma_{\rm sh}$ (the latter equality applying to the
hydrodynamical limit).

Combining this result with Eq.~(\ref{eq:jc4}), one derives a mean
energy per particle downstream of the shock of the order of $e_{\rm
  d}/n_{\rm d}\,\simeq\,\gamma_{\rm rel}$, with $\gamma_{\rm
  rel}\,=\,\gamma_{\rm u}\gamma_{\rm d}\left(1-\beta_{\rm u}\beta_{\rm
  d}\right)\,\simeq\,\gamma_{\rm sh}/\sqrt{2}$ the relative Lorentz factor
between the up- and downstream frames. Superthermal particles share
this mean energy per particle of order $\gamma_{\rm sh}$.

Let us consider now a mean field having some obliquity
$\theta_{B\vert\rm u}$ (measured in the background frame). In the
ideal magnetohydrodynamics (MHD) approximation, the normal ($x-$)
component of the electric field that compensates the electromotive
force $-\boldsymbol{\beta_{\rm u}} \times \boldsymbol{B}$ vanishes and
thus the normal ($x-$) component $B_x$ of the magnetic field remains
unchanged in the shock transition.  In contrast, the transverse
component of the electric field is continuous at the shock front,
therefore $\beta_{\rm u} \boldsymbol{B_{{\rm u},\perp}}\,=\,\beta_{\rm
  d} \boldsymbol{B_{{\rm d},\perp}}$, the $\perp$ symbol denoting the
component transverse to the shock normal; $B_\perp$ is thus amplified
by a factor $w_{\rm d}^{-1}$ when crossing the shock front, in the
shock front frame. However, taking into account the Lorenz factors
involved when transforming from the flow co-moving frames to the front
frame and vice-versa, the global amplification of the transverse
magnetic field component is by the compression ratio $r \equiv u_{\rm
  u}/u_{\rm d} = 2\sqrt{2} \gamma_{\rm sh}$ as for the density.

\subsection{Particle kinematics}\label{sec:dHT}
Because of this tremendous amplification of the transverse component
of the magnetic field in ultra-relativistic shocks, relativistic
shocks generically have a ``superluminal" configuration, which means
that one cannot find a Lorentz transform with a sub-luminal velocity
along the shock surface that would cancel the electric field and
have the flow moving along the magnetic field (the so-called de
Hoffmann Teller frame, see \citet{1950PhRv...80..692D} as well as
\citet{BK90}). Consider indeed a frame moving at the velocity
$\boldsymbol{\beta_{\rm dHT}}$ along the direction of the projection
$\boldsymbol{B_{\rm u,\perp}}$ on the shock surface; then in order
to cancel the electric field, one needs to boost to a frame moving
with a velocity $\boldsymbol{\beta_{\rm dHT}}$ such that
$\boldsymbol{E}+ \boldsymbol{\beta_{\rm dHT}} \times
\boldsymbol{B_{\rm u,\parallel}} = 0$; inserting $\boldsymbol{E} = -
\boldsymbol{\beta_{\rm u}} \times \boldsymbol{B_{\rm u,\perp}}$, one
obtains
\begin{equation}
  \beta_{\rm dHT} = - \beta_{\rm u} \frac{B_{{\rm u},\perp}}{B_{\rm
    u,\parallel}} = \beta_{\rm sh} \gamma_{\rm sh} \tan \theta_{B,\vert\rm u} \ .
\end{equation}
Since $\beta_{\rm dHT}$ must be smaller than one, the condition for
having a Lorenz transform which wipes out the electric field in the
front is very restrictive, because it requires
\begin{equation}
\label{eq:dHT}
\gamma_{\rm sh} \sin \theta_{B,\vert\rm u} \,<\, 1 \ .
\end{equation}
When $\gamma_{\rm sh} \sin \theta_{B,\vert\rm u} > 1$, the
intersection point of a field line on the shock plane moves at a
velocity larger than the velocity of light, hence the word
``super-luminal" for qualifying such a shock. Quite generically,
ultra-relativistic shocks are super-luminal and the matter flow always
crosses the field lines and experiences an electric field.

As for superthermal particles, if they cross the shock front from
downstream to upstream, it means that the $x-$component of their
$3-$velocity $\beta_x$ is positive in the shock front frame, or
equivalently larger than $\beta_{\rm sh}$ in the background plasma
rest frame. Thus, in this frame, the pitch angle $\alpha$ of the
superthermal particle motion with respect to the shock normal is such
that $\sin \alpha < 1/\gamma_{\rm sh}$; the particle distribution of
superthermal particles is extremely anisotropic in the rest frame of
the background plasma, although it can be considered as nearly
isotropic in the shock rest frame.

The trajectory of superthermal particles upstream of the shock is a
crucial point in the theory of relativistic shocks, since its
extension determines the size of the precursor, which corresponds to
that region in which the superthermal and background populations
intermix. To a good approximation, one can consider that a
superthermal particle, while traveling upstream with momentum
$p_{\vert\rm u}$, is caught up by the shock front once its momentum is
deflected by an angle $\gtrsim\,1/\gamma_{\rm sh}$, because the
$x-$component of the particle $3-$velocity then falls below
$\beta_{\rm sh}$ and the shock is trailing right behind the
particle. In the presence of an oblique background magnetic field,
this takes place on a timescale $t_{\rm u\vert u}\,\simeq\,\omega_{\rm
  L,0}^{-1}/\gamma_{\rm sh}$, with $\omega_{\rm L,0}\,\equiv\,e B_{\rm
  \vert u}/p_{\rm \vert u}$ the gyrofrequency in the background
field. Hence the extension of the precursor
\citep[][]{Achterberg2001,MN2006,PLM2009}:
\begin{equation}
  \ell_{\rm prec\rm\vert
  u}\,=\,\left(1-\beta_{\rm sh}\right)t_{\rm u\vert
    u}\,\simeq\,\frac{\omega_{\rm L,0}^{-1}}{2\gamma_{\rm sh}^3}
\end{equation}
As $B_{\rm \vert u}\,\rightarrow\,0$, the size of the precursor becomes
controlled by the scattering of the superthermal particles in the
self-generated turbulence; this case will be discussed further on.

\section{What is a Collisionless Relativistic Shock?}\label{sec:theory}
The concept of collisionless shock relies on the development of
collective effects, including the Landau effect, that control the
dissipation process, i.e. the transformation of the flux of kinetic
energy through the generation of several kinds of entropy. The main
entropy generation is the formation of a thermal distribution of
ultra-relativistic temperature, with (proper) temperature downstream
$T_{\rm d} \,\sim\, \gamma_{\rm sh} m c^2$ (in units in which $k_{\rm
  B}\,=\,1$) or merely the isotropization of the distribution
function, as mentioned previously in the case of ultra-relativistic
shocks.\footnote{Of course shocks do not have to be \ultrarel\ and can
  have any supersonic speed $\lsim c$. This is a critical aspect of
  gamma-ray burst afterglows where the jet shock slows from
  \ultrarel\ to \transrel\ to \nonrel\ speeds. \Transrel\ shocks are
  difficult to describe analytically but can be treated directly with
  PIC or \mc\ techniques.}
This isotropic distribution function can contain a superthermal
component in the form of a quasi-power-law tail; this is what is
expected for the generation of very high energy particles in
astrophysics to account for gamma-ray radiation, for the production of
very high energy cosmic rays, especially ultra-high energy cosmic
rays, or even high energy neutrinos. The generation of a superthermal
tail does not require more energy, of course; for a given global
energy budget, a fraction of the incoming energy is transferred into
the pressure of a cosmic ray component corresponding to a significant
fraction of the thermal pressure:
\begin{equation}
\label{XICR}
P_{\rm cr} = \xi_{\rm cr} \gamma_{\rm sh}^2 n_{\rm u}m c^2 \ .
\end{equation}
Note that the above equation neglects factors of order unity (see in
particular Eq.~\ref{eq:jc4}).  Typically a conversion factor $\xi_{\rm cr}$
of order 10 percent is found in numerical simulations and is expected
to account for the observations.

Such particle acceleration cannot occur without the excitation of some
electromagnetic turbulence that also takes a significant fraction of
the incoming energy:
\begin{equation}
\label{eq:wem}
W_{\rm em} = \xi_B \gamma_{\rm sh}^2 n_{\rm u}m c^2 \ .
\end{equation}
How this turbulent component is generated will be discussed in
Sec.~\ref{sec:prec}.
Of course a preexisting mean magnetic field or background turbulence
may be present in addition to self-generated turbulence.

In the next section we describe two limiting cases: that of a
magnetized shock with a strong mean ambient field, and a weakly
magnetized shock where the self-generated turbulence dominates the
mean field. We then present a general phase diagram for the physics of
these relativistic collisionless shock waves and discuss the
consequences for Fermi acceleration.

\subsection{With a mean field}
Early work on the theory and simulation of a \rel\ shock with a
transverse mean field in pair plasmas, with an obvious application to
the termination shock of Pulsar Wind Nebulae, was worked out by
J. Arons and collaborators in several steps
\citep[][]{1988PhFl...31..839A,Hoshino92,1994ApJ...435..230G}
\citep[see also][]{1984ApJ...283..710K}. In this 1-D simulation, the
incoming pair plasma is partially reflected back at a magnetic
barrier, which is formed as the transverse magnetic field is
compressed in the slow-down of the flow. A fraction of the incoming
plasma is transmitted whereas a significant fraction is reflected
making two opposite loops in phase space that end up in the downstream
flow, producing a synchrotron resonant electromagnetic field that
propagates both forwards and backwards in the form of an extraordinary
mode. The backward mode, because of its compressive nature, heats the
downstream flow.

When the flow is composed of protons and electrons, the intense
coherent forward wave produces, through a ponderomotive force, an
electrostatic wake of large amplitude, that brings the incoming
electrons to a relativistic temperature up to equipartition with the
protons. This eventually quenches the formation of the electrostatic
wake by the ponderomotive force, which is proportional to $q^2/m$,
once the relativistic mass of the electrons reaches that of the
proton. Once electrons have been heated to near equipartition with the
protons, the formation of the shock becomes similar to the case of a
pair plasma.

This scenario depends on a crucial parameter called ``magnetization,"
which is defined as the ratio of the flux of magnetic energy across
the shock to the flux of incoming matter energy:
\begin{equation}
\label{eq:sigma}
\sigma \,\equiv\, \frac{B_{\perp}^2/4\pi}{\gamma_{\rm sh}^2 n_{\rm
    u}mc^2} \ .
\end{equation}
Note that $B_\perp$ is defined in the shock frame, following our
notation convention.  The magnetization parameter is sometimes defined
relative to the incoming kinetic energy $\gamma_{\rm
  sh}\left(\gamma_{\rm sh}-1\right)n_{\rm u}mc^2$, but in the
ultra-relativistic limit $\gamma_{\rm sh}\,\gg\,1$, these two
definitions are equivalent.

Apart from the angular dependence, $\sigma$ depends essentially on the
magnetization of the ambient plasma, since $B_{\perp} = \gamma_{\rm
  sh} B_{\perp\vert\rm u}$. In the typical interstellar medium,
$\sigma \,\sim\, 10^{-9}$, thus a ultra-relativistic shock in the
interstellar medium corresponds to a very low magnetization.

\subsection{Weakly magnetized shocks}
In this subsection, we indicate the requirement for producing
micro-turbulence in a weakly magnetized plasma.
At very low magnetization, a self-consistent process develops such
that a fraction of incoming particles is reflected back into the
upstream region. These reflected particles trigger some form of
streaming instability which generates precursor electromagnetic
turbulence which eventually produces enough scattering to reflect
particles back to the subshock.

Recall indeed from the previous section that particles that outrun the
shock form a beam with opening angle $\,\simeq\,1/\gamma_{\rm sh}$ in
the rest frame of the background plasma. Such a configuration is prone
to beam-plasma instabilities.

The shock thus behaves as a self-sustaining self-generating
structure. Moreover this process can allow for the development of a
superthermal tail, which expands the size of the precursor (which
increases as some power of gyroradius $r_{\rm g}$, hence of particle
energy). In this description, superthermal particles originate from
shock-reflected particles; since both populations share a common mean
energy per particle of $\gamma_{\rm sh}$, we will not distinguish one
from the other in the following and we will use ``returning
particles'' or ``superthermal particles'' interchangeably.

Consider a streaming instability, triggered by the beam of returning
particles, with a growth rate $g_{\rm inst\vert u}$ defined in the
rest frame of the background plasma. In the present situation, an
instability is relevant only if its growth time is shorter than the
time spent by the background plasma in the shock precursor. Assume
also that, although weak, the background magnetic field nevertheless
sets the size of the precursor: $\ell_{\rm prec\vert
  u}\,\sim\,\omega_{\rm L,0}^{-1}/\gamma_{\rm sh}^3$. Therefore a
streaming instability emerges as a good candidate for generating
turbulence in the precursor if its growth rate is such that
\begin{equation}
\label{eq:ginst}
g_{\rm inst\vert u} \,>\,\gamma_{\rm sh}\omega_{\rm c} \ ,
\end{equation}
where $\omega_{\rm c} = eB_{\rm \vert u}/mc$ is the cyclotron frequency
of the background plasma; the above uses $\omega_{\rm
  L,0}\,=\,\omega_{\rm c}/\gamma_{\rm sh}^2$ since $p_{\rm \vert
  u}\,\simeq\,\gamma_{\rm sh}^2 mc$ for a superthermal particle in
the background plasma rest frame. This is a severe constraint at a
relativistic shock; expressed for a proton plasma as a function of
magnetization, this constraint becomes:
\begin{equation}
\label{eq:sigmalim}
\sigma < \gamma_{\rm sh}^{-2}\frac{g_{\rm inst\vert u}^2}{\omega_{\rm pi}^2}
\end{equation}
with $\omega_{\rm pi}\,\equiv\,\left(4\pi n_{\rm
  u}e^2/m_p\right)^{1/2}$.
The process not only requires a very low magnetization, it also
requires instabilities that work on micro-physical scales since the
criterium (\ref{eq:ginst}) excludes slower MHD dynamics.
This latter criterium is the essential reason for which
the phenomenology of ultra-relativistic shocks displaying intense
magnetic fields and superthermal tails, like the termination shocks
of gamma-ray bursts and pulsar wind nebulae, unavoidably requires a
micro-physical description.

\subsection{Classification in term of the magnetization. Phase diagram}
In this subsection, we present the micro-instabilities that are
possible candidates for the generation of the magnetic turbulence.
The maximum growth rates of standard beam-plasma instabilities are
(within a factor of order unity) $(\omega_{\rm pb}^2 \omega_{\rm ep}/
\gamma_{\rm b\vert u}^2)^{1/3}$ for the electrostatic two stream
instability ($\gamma_{\rm b\vert u}$ represents the typical Lorentz
factor of the beam of superthermal particles in the upstream plasma
frame, see below), $(\omega_{\rm pb}^2 \omega_{\rm ep})^{1/3}$ for the
oblique two-stream instability, and $\omega_{\rm pb}$ for the Weibel
instability, where $\omega_{\rm pb}$ is the (relativistic) plasma
frequency of the beam of returning particles:
\begin{equation}
  \omega_{\rm pb}\,=\,\left(\frac{4\pi n_{\rm cr\vert
      u}e^2}{\gamma_{\rm b}m}\right)^{1/2}
\end{equation}
and $\omega_{\rm ep}$ represents the electron plasma frequency of the
background plasma. Using Eq.~(\ref{XICR}) with $n_{\rm cr\vert
  u}\,=\,\gamma_{\rm sh}P_{\rm cr}/\left(\gamma_{\rm sh} mc^2\right)$,
and $\gamma_{\rm b\vert u}\,\simeq\,\gamma_{\rm sh}^2$, one finds
$\omega_{\rm pb} \,\simeq\, \sqrt{\xi_{\rm cr}} \omega_{\rm p}$, where
the conversion parameter $\xi_{\rm cr}$ is defined by relation
(\ref{XICR}) and $\omega_{\rm p}$ is the electron plasma frequency
$\omega_{\rm pe}$ in a pair plasma, or $\omega_{\rm pi}$ in a proton
electron plasma.

Note that a growth time scale of the order of $\omega_{\rm pi}^{-1}$
can be compatible with condition (\ref{eq:ginst}) if the magnetization
is low enough, but not the proton cyclotron time $\omega_{\rm
  ci}^{-1}$. A detailed analysis reveals that the only magnetized
waves which could be excited are the electron whistler waves
\citep{LP2010}.

The two stream instability and the oblique two stream instability grow
through a \v{C}erenkov resonance of the electrostatic (Langmuir) wave
with the background electrons. When the background electrons are
heated to relativistic temperature, the electrostatic waves become
superluminal and thus these instabilities are quenched by resonance
suppression \citep{2009ApJ...699..990B,LP2011,2012ApJ...744..182S}.
Nevertheless, the oblique two stream instability certainly plays an
important role at the beginning of the precursor by heating the
background electrons.

A phase diagram involving the two parameters $\sigma$ and $\gamma_{\rm
  sh}$ can be drawn to delimit the region in which micro-turbulence
can be self-generated. For the most often discussed instability,
namely the filamentation/Weibel instability, the criterion is simple:
$\gamma_{\rm sh}^2 \sigma < \xi_{\rm cr}$. This limit is a first order
approximation to the development of Weibel-like turbulence in the
precursor of relativistic shocks; in the following, we show that a
refined description of the physics of the precursor introduces a
``buffer effect'' which modifies somewhat this limit at moderate
magnetization.

\begin{figure}
\begin{center}
\includegraphics[width=0.8\columnwidth]{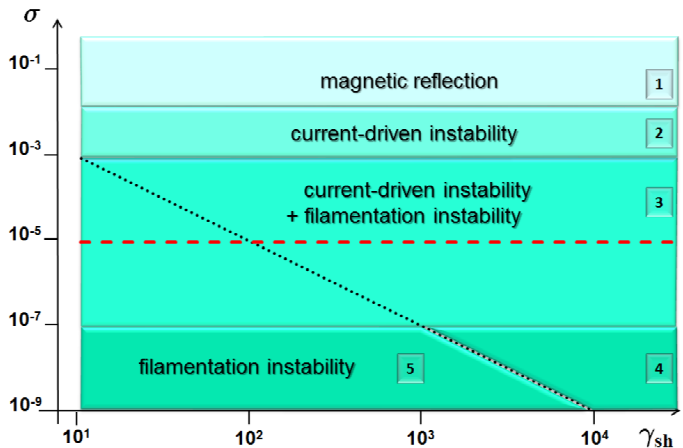}
\caption{Phase diagram of relativistic shocks ($\gamma_{\rm
    sh}\,\geq\,10$) in the plane $(\gamma_{\rm sh},\sigma)$. The
  dotted diagonal line delimits from above the region in which the
  filamentation instability can grow in the background plasma on a
  precursor crossing timescale, in the absence of plasma slow-down. In
  the presence of a background magnetic field, the background plasma
  is slowed down as it enters the precursor (see Sec.~\ref{sec:cdf}),
  hence the filamentation instability can grow in all of region~3. The
  dashed horizontal line delimits from above the region in which the
  scattering in the micro-turbulence is sufficiently strong to allow
  Fermi acceleration, see Sec.~\ref{sec:fermi}.  In region~1, the
  shock forms through the reflection on the compressed background
  magnetic field; at lower magnetizations, the shock forms through the
  build-up of a microturbulent magnetic barrier. The dominant
  instabilities are: filamentation in region~5, filamentation and
  current-driven instability (see Sec.~\ref{sec:cdf}) in region~3, and
  the current driven instability in region~2.}
\label{fig:phased}
\end{center}
\end{figure}

\subsection{Fermi acceleration}\label{sec:fermi}
Similarly to what has been proposed for non-relativistic shocks,
relativistic shocks can develop a superthermal tail thanks to the
Fermi process. Indeed the excitation of magnetic turbulence that
extends in the precursor and is also transmitted to the downstream
flow allows for fast scattering of superthermal particles that can
cross the shock front back and forth and thus gain energy.  However at
relativistic shocks, the Fermi process has some specific features.

We assume that the magnetic disturbances that scatter particles are
almost frozen in the plasma flow both upstream and downstream (this
will turn out to be true with Weibel turbulence). The particle energy
gain through the Fermi process depends on the relative velocity
$\beta_{\rm rel}$ between the upstream and downstream flows. Let a
particle, having a pitch angle cosine $\mu_1$ before scattering, cross
the shock front, be scattered and come back with a pitch angle cosine
$\mu_2$. Its energy gain is
\begin{equation}
\label{eq:egain}
G = \frac{1-\beta_{\rm rel} \mu_1}{1- \beta_{\rm rel} \mu_2}
\ .
\end{equation}
As defined above, $\mu_1$ and $\mu_2$ are expressed in the upstream
rest frame and $\beta_{\rm rel}$ represents the velocity of the
downstream plasma relative to the upstream frame; changing the sign
convention for $\beta_{\rm rel}$, one would obtain the same formula
with $\mu_1$ and $\mu_2$ now both expressed in the downstream rest
frame.
In equation~(\ref{eq:egain}), and for the following discussion, it is
implicitly assumed that the shock is unmodified by Fermi acceleration
with a discontinuous jump between the upstream and downstream regions.

Choosing the downstream reference frame, $\beta_{\rm rel} \simeq 1-
1/\gamma_{\rm sh}^2$ and the pitch angle cosine of a particle crossing
the front towards downstream has $-1 < \mu_1 < \beta_{\rm sh}$; a
particle coming back from downstream to upstream has $\beta_{\rm sh}<
\mu_2 < 1$.

The average energy gain can be calculated easily in the case of
non-relativistic shocks only, because the distribution functions are
almost isotropic and $\mu_1$ and $\mu_2$ independent; one then finds:
\begin{equation}
\langle G\rangle = \left(1+\frac{2}{3}\vert \beta_{\rm rel} \vert\right)^2 \ .
\end{equation}
Moreover, because of the almost isotropic distributions, the
probability of escape through advection in the downstream plasma can
also be easily estimated: $P_{\rm esc} = 4 \left\vert\beta_{\rm
  d}\right\vert$. The index of the power law distribution ${\rm
  d}N/{\rm d}p\,\propto\, p^{-s}$ is then obtained as \citet{Bell78a}:
\begin{equation}
  s \,=\, 1 - \frac{\ln\left(1-P_{\rm esc}\right)}{\ln \langle G\rangle} \simeq
1+ 3 \left\vert\frac{\beta_{\rm d}}{\beta_{\rm rel}}\right\vert \ ,
\end{equation}
i.e. $s = 2$ for the strong adiabatic shock where
$\left\vert\beta_{\rm rel}\right\vert = 3 \left\vert\beta_{\rm
  d}\right\vert$.

For relativistic shocks, the strong anisotropy of the distribution
functions makes these calculations more complicated,
\citep[e.g.][]{EJR90,1999MNRAS.305L...6G,2000ApJ...542..235K,Achterberg2001}
 see however \citet{2003ApJ...591..954V}, \citet{LP2003}, or \citet{KW2005}
 for an alternative point of view. If one were to consider a random
incident pitch angle cosine $\mu_1$, then $\mu_2 \simeq 1 - {\cal
O}( 1/\gamma_{\rm sh}^2)$ would imply $\langle G\rangle \sim
\gamma_{\rm sh}^2$.
However, this maximum limit can only be achieved for particles
crossing from downstream to upstream for the first time.
Indeed, particles participating in a further Fermi cycle are caught up
by the front upstream with $\mu_1 \simeq 1 - {\cal O} (1 /\gamma_{\rm
  sh}^2)$ so that the gain falls to a few only; precisely it is always
close to 2 after the first Fermi cycle
\citep[e.g.][]{Achterberg2001,LP2003}.

\begin{figure}
\begin{center}
\includegraphics[width=0.8\columnwidth]{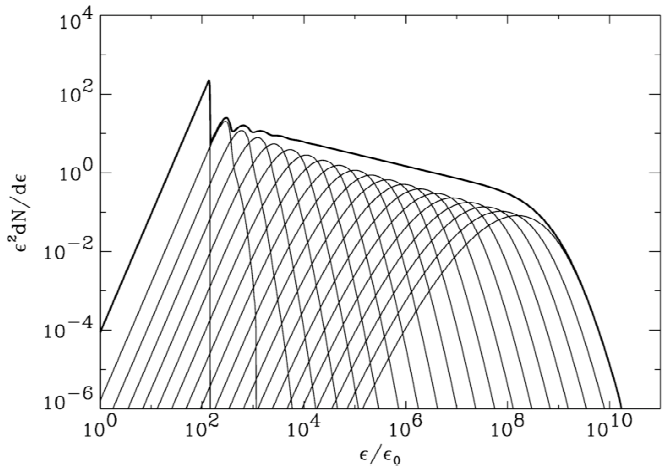}
\caption{Spectrum of superthermal particles as collected far
  downstream (thick line) for $\gamma_{\rm sh}\,=\,100$, obtained from
  a test-particle Monte Carlo simulation, assuming isotropic
  scattering both upstream and downstream of the shock, as in an
  unmagnetized shock \citep{LP2003}. The thin lines indicate the
  populations of particles that have experienced $0,\,1,\,\ldots$
  Fermi cycles through the shock: as the number of cycles increases,
  so does the mean energy of the population, but with a smaller number
  of particles due to the finite $P_{\rm esc}$.}
\label{fig:fermi}
\end{center}
\end{figure}

The index can be defined through the equation
\begin{equation}
\label{eq:srel}
\langle 1-P_{\rm esc}(\mu_2)\rangle \langle G(\mu_1, \mu_2)^{s-1}\rangle = 1  \ ,
\end{equation}
and because $G$ remains close to 2 at each cycle, the index turns out to be close to
\begin{equation}
\label{eq:srel2}
s \,\simeq\, 1 - \frac{\ln \left(1-P_{\rm esc}\right)}{\ln \langle G\rangle} \ .
\end{equation}
The numerical estimate is $1-P_{\rm esc}\,\simeq\, 0.4$ leading to $s
\simeq 2.3$. An illustration of the spectrum of accelerated particles
for an unmagnetized, unmodified shock with $\gamma_{\rm sh}\,=\,100$
is shown in Fig.~\ref{fig:fermi}.

As we describe in Sec.~\ref{MC1}, basic requirements of energy and
momentum conservation mean the test-particle power law shown in
Fig.~\ref{fig:fermi} will be modified to some extent if shocks
accelerate cosmic rays efficiently.  If the scattering mean free path
is an increasing function of particle momentum as expected, the
accelerated spectrum will obtain a concave shape being harder at high
energies before a turnover produced by finite shock effects.

In addition, if the shock is sweeping up \nonrel\ matter, such as
the ISM, the injection of highly anisotropic thermal particles must
be described consistently with CR production. Currently, the only
techniques that can handle thermal particle injection and efficient
non-linear CR production in \rel\ shocks are PIC simulations and \mc\
techniques, described in greater detail in Sec.~\ref{MC1}.

\section{Physics of the precursor}\label{sec:prec}
The main physical processes that develop in the precursor of a
relativistic collisionless shock undergoing Fermi acceleration are the
following:
\newlistroman

\listromanDE The incoming precursor plasma slows before the
  sharp transition at the viscous subshock due to transverse
  currents. This occurs with ordered or disordered magnetic fields for
  both pair and ionic plasmas and produces an important buffer effect
  in the precursor where a highly relativistic shock far upstream
  turns into a mildly relativistic one deep in the precursor, close to
  the shock front.
If Fermi acceleration is efficient the backpressure of accelerated
particles produces an additional shock modification forced by
conservation of energy and momentum.
    
\listromanDE We argue that the Weibel/filamentation instability is the
dominant generator of precursor magnetic turbulence; this instability
does not quench when the electrons are heated to relativistic
temperatures, but it saturates when the beam dispersion becomes too
large.
  
\listromanDE Then we develop the important concept of the Weibel frame
for both pair and ionic plasmas. This is the frame where the
turbulence is purely magnetic and where particles undergo a simple
pitch-angle scattering process.

\listromanDE We discuss the crucial heating process of background
electrons and relate it to the drag force experience by the background
particles.
    
\listromanDE We discuss some aspects of the non-linear theory for the
formation of the filaments.

\listromanDE We then return to the physics of particle scattering in
Weibel turbulence and derive it from the the theory of Hamiltonian
chaos (the details are given in App.~\ref{sec:appA}).

\subsection{Dynamics in the precursor with a mean field; buffer effect}\label{sec:cdf}
The slow-down of the incoming plasma with respect to the shock front
will be analyzed in a dedicated Section~\ref{sec:sdu} for the case of
an unmagnetized plasma. In this section we examine the simpler case of
the slow-down due to a transverse mean field.

\subsubsection{The case of a pair plasma}
Consider first the beam of returning particles. Because of the
transverse mean magnetic field (aligned say along $\boldsymbol{z}$),
electrons and positrons are deflected in opposite directions (along
$\boldsymbol{y}$) during their orbit around $B$ and thus a current
transverse to the shock normal and to the mean field is generated,
with density $j_{\rm b} = n_{\rm cr}ec$ (measured here in the front
frame). One can show that this current is compensated to an excellent
accuracy and on a short timescale by a current carried by the incoming
background plasma, once it enters the precursor. As the density of
this background current density can be written $j_{\rm bg} =
\gamma_{\rm sh}n_{\rm u} e \beta_y c$ in the front frame, with $n_{\rm
  cr}\,=\,\xi_{\rm cr} \gamma_{\rm sh} n_{\rm u}$, one infers a
transverse velocity $\beta_y \,\simeq\, \xi_{\rm cr}$
\citep{LPGP2014,LPGP2014b}.

The main slow-down effect on the incoming plasma is due to the
deviation of background electrons and positrons due to the development
of the compensating current. Indeed, one can show that the total
Lorentz factor of the background plasma is conserved, while the
development of a transverse velocity $\pm \xi_{\rm cr}$ for
electrons/positrons implies that the center-of-mass of the background
plasma moves solely along $\boldsymbol{x}$ with a Lorentz factor:
\begin{equation}
\label{eq:buffer}
\gamma_{\rm cm} = \frac{\gamma_{\rm sh}}{(1+\xi_{\rm cr}^2 \gamma_{\rm
    sh}^2)^{1/2}} \ .
\end{equation}
Thus any shock front of Lorentz factor larger than $1/\xi_{\rm cr}$
produces a deceleration of the incoming plasma such that the center of
mass moves with a Lorentz factor $1/\xi_{\rm cr}$. This is a crucial
buffer effect that transforms any ultra-relativistic shock into a
mildly one in its precursor from the point of view of the growth of
instabilities.

As discussed in \citet{LPGP2014,LPGP2014b}, this modification of the
motion of the center-of-mass indeed modifies the criterion for the
growth of instabilities in the precursor of a shock, due to time
dilation effects. One can show in particular that the Weibel
instability can now grow provided $\sigma < \xi_{\rm cr}^3$,
independently of the shock Lorentz factor, as illustrated in
Fig.~\ref{fig:phased}. Moreover, the above configuration of a
compensated transverse current turns out to be unstable on a short
timescale, leading to the growth of a current-driven filamentation
instability with growth rate $\sim\,\omega_{\rm pe}$, i.e. faster
than the Weibel/filamentation rate. This latter instability can thus
grow provided $\sigma < \xi_{\rm cr}^2$ at any value of the shock
Lorentz factor, thereby bridging the gap between weakly and
moderately/strongly magnetized shock waves.

\subsubsection{The case of an electron-ion shock}
We consider first an electron-ion plasma where electrons have been
rapidly heated to a relativistic temperature of order $m_pc^2$ (this
heating will be explored in a next section). Their inertia is thus
comparable to that of the protons, so that a significant fraction of
the electrons is reflected back on the shock front along with protons,
leading to a charge neutral beam of returning particles. In such a
case, the situation is similar to the case of a pair plasma discussed
above. If the electrons are relativistically hot, but below
equipartition with the protons, i.e. $T_e = \xi_{\rm th} m_pc^2$ with
$\xi_{\rm th} > m_e/m_p$, one can recover similar results, as will be
presented elsewhere.

If the population of incoming electrons remains cold because the
precursor is too short, then electrons are not reflected back and the
beam of returning current carries a net electric charge $\rho_{\rm cr}
= \xi_{\rm cr} \gamma_{\rm sh} n_{\rm u} e$ (front frame). As the
background plasma compensates this electric charge, a force density
develops along the y-direction, transverse to the shock normal and to
the mean field $B$; in the relativistic MHD approximation, this force
density reads:
\begin{equation}
\label{eq:sd-force}
f_y \,=\,-\rho_{\rm cr} E_y \,\sim\, \beta_{\rm sh} \xi_{\rm cr}n_{\rm
  u}\gamma_{\rm sh} eB \ .
\end{equation}
The above equation uses a zero current density along $x$, which is
guaranteed by the assumption of one-dimensional stationary dynamics in
the shock front frame. The work of this force over the length of a
Larmor gyration is $\xi_{\rm cr} \beta_{\rm sh} \gamma_{\rm sh}^2 n_{\rm
  u}m_pc^2$. Thus protons are deviated in the transverse direction
with a velocity $\xi_{\rm cr}$ and the bulk motion is diminished as
previously.

\subsection{Weibel turbulence}\label{sec:weibel}
Because the mean magnetic field must be very weak and the growth of
the instability faster than the cyclotron frequency of electrons in a
pair plasma, or of protons in a protonic plasma, the expected
turbulence must be excited by a fast micro-instability on short length
scales. The typical scale must be at most the inertial scale of
protons: $\delta_i \equiv c/\omega_{\rm pi}$; MHD develops at larger
scales than this inertial scale.

\begin{figure}
\begin{center}
\includegraphics[width=0.7\columnwidth]{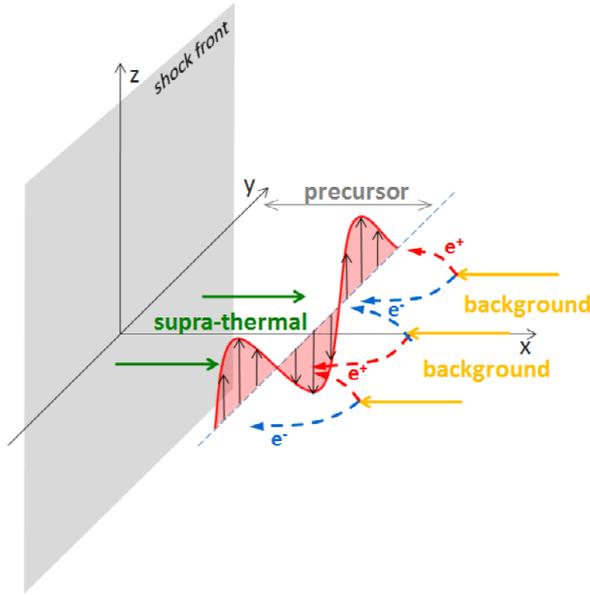}
\caption{Development of the Weibel/filamentation instability in the
  precursor of a relativistic shock: as the background plasma enters
  the precursor defined by the region in which superthermal particles
  exist, the interpenetration of the two populations develops a
  filamentation instability: a transverse magnetic fluctuation of the
  background plasma leads to charge separation in this population,
  forming current filaments at the nodes of the fluctuation, which
  then feeds back positively on the fluctuation. See text for
  details.}
\label{fig:weibel}
\end{center}
\end{figure}

The Weibel instability, which oversteps the other streaming
instabilities and survives to electron heating, is mostly an
electromagnetic instability with a very low phase velocity, and thus
produces slowly propagating magnetic turbulence, quite suitable for
particle scattering and thus for a Fermi process. The mechanism of
the instability is simple: the beam of returning particles is
supposed to be charge neutralized, although when a transverse
magnetic wave has been produced, its quiver force separates electric
charges; this generates an intense electric current, which in turn
amplifies the primary magnetic wave; then it blows up
\citep[see
e.g.][]{1999ApJ...526..697M,2004A&A...428..365W,2006ApJ...647.1250L,
MN2006, 2007A&A...475....1A,2007A&A...475...19A}. Its growth rate is
fast, of the order of $\omega_{\rm pb}$, with respect to the
background frame, and the instability is purely growing in this
frame with a weak electrostatic component. A sketch of the mechanism
of this instability in the precursor of a relativistic shock is
presented in Fig.~\ref{fig:weibel}.

We already mentioned that relativistically hot electrons do not quench
this Weibel instability (to be discussed shortly); however a large
angular dispersion of the beam can stabilize the Weibel instability,
as discussed e.g. in \citet{2011ApJ...736..157R,LP2011}.
A simple description of that saturation effect by the beam dispersion
consists in treating the transverse dispersion of the beam as a
temperature effect; this introduces an effective sound velocity
$\beta_*c \sim c/\gamma_{\rm sh}$, see Eq.~(\ref{eq:wg3}), which
modifies the hydrodynamic response of the beam fluid to the quiver
force that separates opposite charges. The instability then disappears
(in this fluid description) when $\Im\omega \,\lesssim\,
\beta_*\omega_{\rm p}$, i.e.  when $\gamma_{\rm sh}$ becomes smaller
than $1/\xi_{\rm cr}^{1/2}$. A more detailed analysis with some
consequences of this saturation effect will be addressed in a
forthcoming paper.

\subsection{The Weibel frame in a pair plasma}
For the estimation of the efficiency of scattering and thus of the
Fermi process, and also for estimating the radiation efficiency of an
ultra-relativistic shock, it is very important to know what is the
proper frame of Weibel turbulence.

The Weibel frame is defined as the frame where the electrostatic
potential vanishes. Its determination is detailed in
App.~\ref{sec:appA}, where it is shown that it propagates at the speed
$\xi_{\rm cr}$ with respect to the background plasma. This has two
important consequences, namely that the high energy returning
particles are scattered with respect to that frame and that the
background plasma suffers a drag force which slow it down and heats
the electrons.
The Weibel frame is the suitable frame to calculate the growth rate
of the instability, which is purely magnetic and purely growing, since
the charge separation in the beam is exactly compensated by an
opposite charge separation in the background plasma.

In the front frame, a strong electrostatic potential $\delta\Phi$ is
set-up; its magnitude can be determined by the Lorentz transform from
the Weibel frame to the front frame: $\delta\Phi \,=\, \gamma_{\rm
  w}\beta_{\rm w} \delta A_{x\vert\rm w}$, defining $\beta_{\rm w}$
(resp. $\gamma_{\rm w}$) as the velocity (resp. Lorentz factor) of the
frame in which $\delta\Phi$ vanishes, relatively to the shock front
frame. Similarly, the $x-$component of the magnetic potential reads
$\delta A_x \,=\, \gamma_{\rm w} \delta A_{x\vert\rm w}$, so that
$\delta\Phi \,=\, \beta_{\rm w} \delta A_{x} \,\simeq\, -\beta_{\rm
  sh} \delta A_x$. In the front frame, there must be a net total
electric charge density fluctuation $\delta\rho$, also determined
through a Lorentz transform: $\delta\rho = \gamma_{\rm w}\beta_{\rm w}
\delta j_{x \vert\rm w}/c$; the net current density fluctuation is
$\delta j_x = \gamma_{\rm w} \delta j_{x \mid\rm w}$ so that
$\delta\rho \,\simeq\, -\beta_{\rm sh} \delta j_x/c$. It is assumed
here that in the Weibel frame, the total charge density perturbation
$\delta\rho_{\vert\rm w}$ vanishes. The Maxwell equation that
determines the electromagnetic field as a function of the current
density, namely $\Delta \delta A_x = -4\pi j_x/c$, is then the same as
the Poisson relation between the electrostatic potential and the
electric charge.

To conclude this section on the Weibel instability in a pair shock,
let us show that it is quite convenient to calculate the growth rate
of the Weibel instability in the Weibel frame. Indeed in this frame
$\delta\Phi_{\rm \vert w} = 0$ and $\delta \rho_{\rm tot\vert w} = 0$,
which allows one to write two simple coupled equations to derive the
instability growth rate in the cold approximation. Switching over to
Fourier space, assuming $k_x\,=\,0$ as before, we insert $\delta
\rho_{\rm b\vert w} = - \delta \rho_{\rm bg\vert w}$ and $\beta_{\rm
  bg\vert w} = - \xi_{\rm cr}$ in the dynamical equation
$\partial_\alpha\partial^\alpha\delta A^\mu\,=\,-4\pi\delta j^\mu$ and
in the background dynamical equation, to obtain:
\begin{eqnarray}
(-\omega^2+\omega_{\rm p}^2 + k_\perp^2 c^2) \delta A_{x\vert\rm w} & \,\simeq\, & - 4\pi
  \beta_{\rm b\vert w} c^2 \delta \rho_{\rm bg\vert w} \nonumber
  \\ \omega^2 \delta \rho_{\rm bg\vert w} & = & \xi_{\rm cr}
  \frac{\omega_{\rm p}^2}{4\pi} k_\perp^2 \delta A_{x\vert\rm w}
\ .
\end{eqnarray}
The plasma frequency of the background plasma emerges in the first
equation because of the response of the conduction current to the
electromagnetic perturbation. In details:
\begin{equation}
  \delta j_x\,=\,\sum_\alpha \rho_\alpha\delta \beta_{x\alpha} + \beta_{\alpha}\delta \rho_\alpha
\end{equation}
and the conduction current density perturbation
$\rho_\alpha\delta\beta_{x\alpha}$ is obtained through the dynamical
equation for $\delta \beta_{x\alpha}$:
\begin{equation}
  \rho_\alpha\delta\beta_{x\alpha}\,=\,-\frac{\omega_{p\alpha}^2}{4\pi\gamma_\alpha^2}\delta A_x
\ .
\end{equation}
The strong dependence on $\gamma_\alpha$ justifies the neglect of the
conduction response of the beam of superthermal particles;
furthermore, in the Weibel frame, $\gamma_{\rm bg\vert w}\,\simeq\,1$.

One thus derives the dispersion relation to leading order in
$\omega/kc$:
\begin{equation}
\label{eq:wg}
\omega^2 = - \xi_{\rm cr} \beta_{\rm b\vert w} \omega_{\rm p}^2
\frac{k_{\perp}^2c^2}{\omega_{\rm p}^2 + k_{\perp}^2c^2} \ .
\end{equation}

This is a purely growing magnetic mode. If we take into account the
case $k_x \neq 0$, two modifications are introduced: a complete
Laplacian operator in the Maxwell equation and a Doppler effect in the
dynamical equation of the background plasma, with $\partial_t^2
\mapsto (\partial_t - \xi_{\rm cr} c \partial_x)^2$ which leads to a
simple modification of the dispersion equation:
\begin{equation}
\label{eq:wg2}
(\omega + \xi_{\rm cr} k_xc)^2 = - \xi_{\rm cr} \beta_{\rm b\vert w}
\omega_{\rm p}^2 \frac{k_{\perp}^2c^2}{\omega_{\rm p}^2 + k^2c^2} \ .
\end{equation}
The growth rate is invariant by parity; thus in the Weibel frame,
forward and backward waves propagate with a frequency $\xi_{\rm cr}
k_xc$ with the same growth rate. A strong Doppler effect is produced
in the front frame, so that both forward and backward waves are caught
up by the shock front.

\subsection{The Weibel instability in an electron-ion shock}
We now assume a plasma made of protons and hot electrons and extend
the calculations to include the Landau effects in the electromagnetic
response. The detailed calculations are presented in
App.~\ref{sec:appB}.

As for the pair shock, one can derive the velocity of the Weibel frame
by analyzing the generation of electric charge by the quiver force,
taking into account the susceptibilities of the various plasma
components. To circumvent the technical difficulty associated with the
Lorentz transform of these susceptibilities from one frame to another,
it is more convenient to rely on a Lorentz covariant formulation of
linear response theory, as formulated by \citet{1986islp.book.....M}
and \citet{2007A&A...475....1A}, which involves the polarization
tensor of species $s$ $\alpha^{\mu}_{s,\nu}$ defined through $\delta
j_s^{\mu} = \alpha^{\mu}_{s,\nu} \delta A^{\nu}$. The combination of
Maxwell equations in the Lorentz gauge, namely
$\partial_\mu\partial^\mu \delta A^\nu\,=\,-\sum_s \delta j_s^\nu$ and
of the above response then leads to the definition of the Weibel frame
as that in which
\begin{equation}
  \sum_s\,\alpha^{01}_s\,=\,0
\ .
\end{equation}
The calculations of $\alpha^{01}_s$ in a generic frame, for streaming
hot electrons, are too lengthy to be reported here; these calculations
will be presented elsewhere. Nevertheless, one
obtains the same result as before, $\beta_{\rm w\vert
  u}\,\sim\,\xi_{\rm cr}$.

The concept of Weibel frame is essential in the prospect of particle
acceleration and radiation, since this is the frame in which particles
are purely scattered (at least almost, for there is an induction
electric field significantly smaller than the magnetic field). The
conclusion of this section is that particles are scattered by the
magnetic field of Weibel turbulence with respect to a frame moving
forward at sub-relativistic speed ($\sim\,\xi_{\rm cr}$) with respect
to the background plasma.

\subsection{Electron heating and the drag force}\label{sec:sdu}
Electrons of the background plasma can be heated through two
possible electric fields generated by the Weibel instability: one is
the strong motional electrostatic field measured in the front frame
due to the Lorenz transform, oriented both perpendicular to the
magnetic field fluctuation and to the shock normal; the other is the
induction electric field, which derives from the growth of the
magnetic field energy density, and which is significant during the
stage of instability growth.

The electrostatic field $\delta E_\perp$ is associated with a
potential fluctuation which has been obtained in the front frame
through a Lorentz transform: $\delta\Phi \,=\, \beta_{\rm w} \delta
A_{x}$. The heating due to this field is caused by the scattering of
electrons in the Weibel frame, which leads to a kind of second order
Fermi process in the front frame, because a variation of $x-$momentum
in the Weibel frame produces a variation of energy in the front
frame. This is the strongest electric field when measured in the front
frame with $\delta E_\perp \sim \delta B$.

In the Weibel frame, the inductive electric field derives from:
\begin{equation}
  \delta E_{x\vert\rm w} \,\equiv\, -\frac{1}{c} \partial_t \delta A_{x\vert\rm w} \,\sim\,
  \frac{g_{\rm inst\vert w}}{\omega_{\rm pi}} B_{\vert\rm w} \ .
\end{equation}
Therefore $\delta E_{x}^2 \,\sim\, \xi_{\rm cr} \delta E_\perp^2$ in
the front frame and $\xi_{\rm cr}\,<\,1$; this inductive electric
field nonetheless provides an efficient source of
heating~\citep{2012EL.....9735002G}.\\

\subsubsection{Scattering and heating in the electrostatic field}
In a first approximation, background electrons are scattered by a
static magnetic field in the Weibel frame that moves at speed
$\beta_{\rm w\vert u} \,\simeq\, \xi_{\rm cr}$ with respect to the
background frame. Therefore, in the background frame, their
temperature increases and they suffer a drag force. Because their
energy does not change during scattering in the Weibel frame (as long
as one neglects the induction field effect), they suffer a
modification of their $x-$momentum $\Delta p_{x\vert\rm u}$ and energy
$\Delta \epsilon_{\vert\rm u}$ in the background frame.

Together with $\Delta \epsilon_{\vert\rm w}\,=\,0$, the Lorentz
transform $\Delta \epsilon_{\vert\rm w} = \gamma_{\rm w\vert u}(\Delta
\epsilon_{\vert\rm u} - \beta_w c \Delta p_{x\vert\rm u})$ provides an
essential relation between energy gain and momentum variation:
\begin{equation}
\label{eq:sc1}
\Delta \epsilon_{\vert\rm u} = \beta_{\rm w\vert u} c \Delta p_{x\vert\rm u} \ .
\end{equation}
The stochastic force $\Delta p_{x\vert\rm u}/\Delta t_{\vert\rm u}$,
after statistical averaging, leads to a macroscopic drag force $F_{\rm
  d\vert u}$ and the distribution of background electrons, isotropic
in its rest-frame, gets heated at a rate (background frame):
\begin{equation}
\label{eq:sc2}
k\dot T = \frac{1}{3} \beta_{\rm w\vert u} c F_{\rm d\vert u} \ .
\end{equation}
The drag force is best evaluated in the Weibel frame through a
Fokker-Planck term:
\begin{equation}
  F_{\rm d\vert w}\,=\,\frac{1}{n}\frac{\partial \left(n\langle\Delta
    p_{x\vert\rm w}\rangle\right)}{\partial t}\,=\,\int {\rm d}^3
      p_{\vert\rm w}\, p_{x\vert\rm w}\frac{\partial}{\partial
        p_{i\vert\rm w}}D_{ij}\frac{\partial}{\partial
        p_{j\vert\rm w}}f(\boldsymbol{p_{\rm\vert w}})
\end{equation}
with a momentum diffusion coefficient of the form
\begin{equation}
  D_{ij}\,=\,e^2 <\delta B_\perp^2> \tau_{\rm c\vert w} h_{ij}^\perp
\end{equation}
with $h_{ij}^\perp\,=\,\delta_{ij}-p_{i\vert\rm w}p_{j\vert\rm
  w}/p_{\rm\vert w}^2$ the projector orthogonal to momentum. The
timescale $\tau_{\rm c\vert w}$ represents the coherence timescale of
the random force associated to the Weibel turbulence. Using the Lorentz
invariance of $f$ and the fact that $f$ depends only on
$\epsilon_{\rm\vert u}/T$ in the background frame, one can derive
\begin{eqnarray}
F_{\rm d\vert w} & \,\sim\, & \frac{\beta_{\rm w\vert u} c}{kT} e^2
<\delta B_\perp^2> \tau_{\rm c\vert w}
\end{eqnarray}
up to a prefactor of order unity.

Using the above, we can estimate the temperature that can be achieved
and the slow-down of the background flow that ensues. As discussed
further below -- see Eq.~\ref{eq:bs2} in particular-- the precursor
length is approximately $\delta_i/\xi_B$, as measured in the
background rest-frame, and therefore the background plasma experiences
the drag force for a duration $\xi_B^{-1} \omega_{pi}^{-1}$. The
temperature achieved is then roughly of order
\begin{equation}
\label{eq:sc4}
\xi_{\rm th} \,\sim\, \left(\frac{\xi_{\rm cr}^2 e^2
<\delta B_\perp^2> \tau_{\rm c\vert w} c^2}{m_p^2c^4 \xi_B
  \omega_{\rm pi}}\right)^{1/2} \,\sim\, \xi_{\rm cr} \,\left(\tau_{\rm c\vert w}\omega_{\rm pi}\right)^{1/2}\ .
\end{equation}
In the background frame, the dragging effect is seen as a force that
pushes the plasma forward, with an intensity $F_{\rm d\vert u}=F_{\rm
  d\vert w}$. Over the precursor crossing timescale, the upstream
plasma is displaced in energy by
\begin{equation}
\label{eq:sc5}
\Delta\gamma \,\simeq\, \frac{F_d \delta_i}{m_pc \xi_B} \,\sim\,
\left(\tau_{\rm c\vert w}\omega_{\rm pi}\right)^{1/2}
\end{equation}
indicating that in the front frame, the variation in momentum is of
order unity if $\tau_{\rm c\vert w}\omega_{\rm pi}\,\sim\,1$.

Therefore this scattering process acting on the background plasma is
able to slow-down the whole incoming plasma, with possible
consequences regarding the saturation of the Weibel instability in
regards of the angular dispersion of the beam.

A more complete calculation should take into account the spatial
profiles of the various quantities in a dynamical system; however the
former estimates provide useful orders of magnitude regarding heating
and slow down.

\subsubsection{Heating in the inductive electric field}
The induction field acts during the phase of instability growth and
its heating effect can be estimated with quasi-linear theory which
leads to (in the Weibel frame)
\begin{equation}
\label{eq:ih1}
n\dot T \,\simeq\, \int{\rm d}^3p_{\vert\rm w}\,\, \epsilon_{\vert\rm w}
\frac{\partial}{\partial p_{x\vert\rm w}} \Gamma_{xx}
\frac{\partial}{\partial p_{x\vert\rm w}} f(\boldsymbol{p_{x\vert\rm w}})
\end{equation}
with $\Gamma_{xx} \,\simeq\, (g_{\rm inst\vert w}\omega_{\rm pi})^2
e^2 <\delta B_\perp^2> \tau_{\rm c\vert w}$. The heating rate is thus
a factor $\omega_{\rm pi}^{-1}g_{\rm inst\vert w}/\xi_{\rm cr}$ times that in
the stochastic electrostatic field; in the cold plasma limit, both are
comparable because $g_{\rm inst\vert w}\,\sim\,\xi_{\rm cr}\omega_{\rm pi}$,
but at larger temperatures, heating in the inductive field is slightly
larger. However, this heating process in the inductive field only
applies to the linear growth phase; it is much reduced in the
non-linear phase. Overall, one thus expects to reach a scaling given
by Eq.~(\ref{eq:sc4}).

One can also estimate the electron temperature from the phenomenology
of the relativistic regime of oscillations in very intense fields:
$T_e \sim eB\ell_{\rm c}$. The typical transverse scale of turbulence
is the electron Debye length: $\ell_{\rm c} = \lambda_{\rm D} =
\xi_{\rm th}^{1/2} \delta_i$ in the hot electron regime. Then we get
another estimate compatible with the previous one:
\begin{equation}
\xi_{\rm th} \sim \xi_B \, \, {\rm or} \, \, nT \sim \frac{B^2}{4\pi} \ .
\end{equation}
This is a remarkable result that suggests that the thermal energy
density follows the magnetic energy density and that confinement
structures can occur in the background plasma.

\subsubsection{Weibel filaments}
The above relation suggests that heating of the background plasma
ensures that the thermal pressure inside the filaments counterbalance
the magnetic stress. It is important here to note that the filaments
are built out of the background plasma, not by filamentation of the
beam of returning particles, as is often thought. Indeed, the
transverse momentum of these particles in the background plasma frame
is of the order of $\gamma_{\rm sh}mc$, which is too large to allow
them to be confined in structures of transverse size $\delta_i$, since
$e\delta B_{\vert\rm u}\delta_i \,\sim\, \xi_B^{1/2} mc^2$.

The filaments observed in PIC simulations are thus real, but they form
in the background plasma.  This is tightly related to the fact that
the background plasma moves at a velocity $-\beta_{\rm w\vert u}$ with
respect to the magnetic field.  These filaments are some sort of
coherent structures of a quasi bi-dimensional turbulence, where
inverse cascade develops.

One can derive the general structure of these filaments in the
transverse plane by considering the inhomogeneous equilibrium of the
background plasma in the presence of a finite $A_x$ component
(background plasma frame). Using a non-relativistic description,
appropriate to the transverse dynamics, one finds from the force
balance in the transverse direction:
\begin{equation}
\label{eq:f1}
\nabla_{\perp} \left\{\frac{1}{2} m v^2 + T \ln\left(\frac{n_{\pm}}{\bar n}\right) +q\xi_{\rm cr} A_x\right\} = 0 \ .
\end{equation}
The possibly residual kinetic energy $E_K =  m v^2/2$ is
in the form of irrotational motions, the Lorenz driving being
irrotational.  One obtains Boltzmann type equilibria, assuming the
same temperature for both populations of particles:
\begin{equation}
\label{eq:f2}
n \,=\, \bar n \exp\left[-(E_K + \xi_{\rm cr} qA_x)/T\right] \ .
\end{equation}
In a pair plasma, one gets a charge separation such that
\begin{equation}
\label{eq:f3}
(n_+-n_-)e \,=\, -2n_0e \exp\left(-E_K/T\right) \sinh (\xi_{\rm cr} eA_x/T) \ .
\end{equation}
Because in motion of velocity $-\xi_{\rm cr}$ with respect to the
Weibel frame, its associated current generates a magnetic field such
that:
\begin{equation}
\label{eq:f4}
\Delta_{\perp} A_x = -8\pi \bar n e \xi_{\rm cr}
\exp\left(-E_K/T\right) \sinh \left(\xi_{\rm cr} eA_x/T\right) \ .
\end{equation}
Defining $a\,\equiv\, \xi_{\rm cr} eA_x/T$, we obtain a sinh-Poisson
equation, whose analytical solutions are known:
\begin{equation}
\label{eq:f5}
\Delta_{\perp} a = -\frac{\xi_{\rm cr}^2}{\lambda_{\rm D}^2} \exp\left(-E_K/T\right) \sinh a \ .
\end{equation}
The solutions of that equation are known to be in the form of a double
lattice of filaments of alternating polarity. The size of the
filaments appears clearly as $\lambda = \lambda_{\rm D}/\xi_{\rm cr} =
\sqrt{\xi_{\rm th}}/\xi_{\rm cr} \delta_i$. This estimate does not account
for the kinetic energy contribution that increases the Debye
length. However its contribution is expected to remain not larger than
$\xi_{\rm cr} eB\lambda \,\sim\, \vert a \vert T$, which is also the
estimate of the achieved temperature. In a pair plasma that energy is
sub-relativistic.

In the case of an electron-ion plasma, the solution are similar if the
kinetic energy is negligible. Otherwise, it breaks the symmetry by
giving more importance to the electron contribution. In that case
instead of a sinh-Poisson equation, the filaments are described by a
Poincar\'e equation, whose solutions are of Bennett equilibrium type
with all the filaments having the same polarity and thus undergoing
more merging than in the sinh-Poisson case:
\begin{equation}
\label{eq:f6}
\Delta_{\perp} a = -\frac{\xi_{\rm cr}^2}{\lambda_{\rm D}^2} e^a \ .
\end{equation}

\subsection{Scattering off Weibel turbulence}
The size of the precursor is determined by the residence time of
returning particles. As mentioned earlier, and as seen from the
background frame, they are outrun by the shock front when their pitch
angle has opened up to $1/\gamma_s$ due to a scattering at a frequency
$\nu_s$ such that:
\begin{equation}
\label{eq:bs1}
\nu_{\rm s} \,\sim\, \frac{e^2\langle\delta B_{\rm \vert u}^2\rangle\tau_{\rm
    c\vert u}}{\gamma_{\rm b\vert u}^2 m^2c^2} \ .
\end{equation}
The Lorentz factor of beam particles in the background plasma rest
frame $\gamma_{\rm b\vert u}\,\simeq\,\gamma_{\rm sh}^2$. The above
estimate derives from a random phase approximation
\citep[][]{MN2006,PLM2009} and turns out to be valid even with a more
elaborated method based on Hamiltonian chaos, described in
App.~\ref{sec:appC}.

Measured from the background frame one obtains:
\begin{equation}
\label{eq:bs2}
\ell_{\rm prec\vert u} \,\sim\, \frac{c}{\gamma_{\rm sh}^2 \nu_{\rm
    s}}(1-\beta_{\rm sh}) \,\sim\, \omega_{\rm pi}^{-1}\xi_B^{-1}
    \end{equation}
assuming that $\tau_{\rm c\vert u}\,\sim\,\omega_{\rm pi}^{-1}$. Thus
the travel time of the background plasma across the precursor is about
$(\xi_B \omega_{\rm pi})^{-1}$, which is longer than the growth time of
the instability $(\sqrt{\xi_{\rm cr}} \omega_{\rm pi})^{-1}$, while the
residence time of superthermal particles of Lorentz factor
$\gamma_{\rm\vert u}$ in the precursor is of the order of
\begin{equation}
 t_{\rm res\vert u}\,\sim\, 2\gamma_{\rm sh}^2\ell_{\rm prec\vert u}\left(\frac{\gamma_{\rm\vert
   u}}{\gamma_{\rm b\vert u}}\right)^2\label{eq:tres}
\end{equation}
This residence directly controls the acceleration timescale, which
thus scales as the energy squared of the superthermal particle.
This result has been confirmed by PIC simulations
\citep[][]{SSA2013}.

The value of the coherence time for the superthermal particles is
not a trivial issue, as mentioned in \citet{2007A&A...475...19A}.
Indeed, in an idealized Weibel turbulence with $k_x\,=\,0$, the
conjugate momentum $p_{x} - eA_{x}/c$ is a conserved quantity of the
particle trajectories. Since $p_{x}\,\sim\,\gamma_{\rm sh} mc$ while
$eA_{x}/c\,\sim\,\xi_B^{1/2}\gamma_{\rm sh}mc$ in the front frame,
the former largely dominates, hence superthermal particles would
never be able to go back to the shock if the turbulence were exactly
$x-$independent. Consequently, the finite $k_x$ must play a key role
in the transport of superthermal particles.

The general scaling of the residence time with energy $\epsilon$,
hence of the acceleration timescale $t_{\rm
  acc}\,\propto\,\epsilon^2$, has been obtained in the framework of a
random phase approximation model in \citet{PLM2009}.
Nevertheless, one can also derive this scaling using a model of
Hamiltonian chaos applied to the trajectories of superthermal
particles. This leads to a validation of the scattering law in a more
general situation than the random phase approximation.  In
App.~\ref{sec:appC}, we present a simple model of such chaotic
dynamics; a more complete model will be presented in a forthcoming
publication.

\section{Phenomenological consequences}\label{sec:phenom}
\subsection{Scattering efficiency and maximal energy}
\subsubsection{Onset of acceleration}
As discussed in Sec.~\ref{sec:dHT}, an oblique background magnetic
field in the upstream becomes essentially transverse in the
downstream flow, leading to a super-luminal shock configuration. As
discussed in \citet{2006ApJ...645L.129L} and \citet{PLM2009}, such a
configuration inhibits Fermi acceleration unless the scattering
frequency in the turbulence exceeds the gyrofrequency in the
background field (which controls the rate at which particles are
advected in the downstream plasma). In \citet{2006ApJ...645L.129L},
it was actually speculated that the development of Weibel
micro-turbulence should lead to an efficient Fermi process and this
point of view has been confirmed by PIC simulations.

For scattering in micro-turbulence in the downstream flow, the above
condition for the development of Fermi cycles can be written:
\begin{equation}
\label{eq:acc1}
\frac{e^2 \delta B_{\rm \mid d}^2 \tau_{\rm c \mid d}}{m^2c^2 \gamma^2} \,>\, \frac{eB_{\rm \mid d}}{mc \gamma} \ ,
\end{equation}
with $\tau_{\rm c \mid d} \sim \omega_{\rm pi}^{-1}$. Here $\gamma$
denotes the Lorentz factor of the accelerated particle in the
downstream frame; for the first Fermi cycle,
$\gamma\,\sim\,\gamma_{\rm sh}$. The above can be written in terms of
the parameters $\sigma$ and $\xi_B$ as
\begin{equation}
\label{eq:acc2}
\sqrt{\sigma} \,<\, \xi_B  \ .
\end{equation}
It thus indicates that only weakly magnetized shock waves should be
able to accelerate particles efficiently. For typical values
$\xi_B\,\sim\,10^{-2}$ in the vicinity of the shock front, the above
bound suggests that particle acceleration should take place in
shocks with $\sigma\,\lesssim\,10^{-4}$. This point of view,
developed in \citet{2006ApJ...645L.129L} and \citet{PLM2009} (see
also \citet{2006ApJ...650.1020N} for numerical simulations) has been
beautifully confirmed by PIC simulations, e.g. \citet{SSA2013}.

\subsection{Maximal energies}
The maximal energy that can be reached by the Fermi process in a shock
of a given size and age depends on several factors, in particular the
scattering rate [Eq.~(\ref{eq:bs1})], the energy loss or escape
timescale, and the background magnetic field. The above discussion
indicates that in the front frame, the upstream residence time
typically dominates over the downstream residence time; this effect
arises because the upstream fluctuations move at high velocity
relative to the front, while the downstream perturbations are
essentially static and isotropic in the shock front frame. However, as
discussed in detail by \citet{PPL2013}, the inclusion of a small but
finite background magnetic field changes the scale of the precursor.

To simplify the discussion, one can use the downstream residence time
as a proxy for the acceleration timescale; one would obtain similar
results for a more careful calculation taking into account the effect
of a background magnetic field comparable to that of the interstellar
medium and $\gamma_{\rm sh}\,\sim\,100$, representative of gamma-ray
bursts external shocks.

One can then show that the maximal proton energy is of order
$10^{16}\,$eV at the external shock of a gamma-ray bursts. In
contrast, the maximum electron energy is limited by synchrotron
energy losses to values of the order of $10\,$TeV, independent of
the magnetic field and shock Lorentz factor
\citep{KR2010,PPL2013,2013ApJ...771L..33W}. This can give rise to
GeV synchrotron photons at early times $\sim\,10^2-10^3\,$s, as
observed by the Fermi satellite in some gamma-ray bursts.

\subsection{Radiative diagnostics}
Weakly magnetized shocks, which are prone to Fermi acceleration, are
those in which intense micro-turbulence can be excited in the shock
precursor by streaming instabilities. The fate of this turbulence
downstream of the shock, on timescales $\gg\,\omega_{\rm pi}^{-1}$ has
been the subject of intense debate. The linear dissipation rate of
microturbulence is $\gamma_k\,\simeq\,k^3c^3/\omega_{\rm pi}^2$ in a
relativistic plasma \citep[e.g.][]{2008ApJ...674..378C}, a result
which extends into the mildly non-linear
regime~\citep{2015JPlPh..81a4501L}, suggesting that turbulence on
scales $\sim\,\omega_{\rm pi}^{-1}$ is dissipated on $\omega_{\rm
  pi}^{-1}$ timescales. In contrast, the typical (comoving) size of
the blast of a gamma-ray burst is of the order of $R/\gamma_{\rm
  sh}\,\sim\,10^7\,c/\omega_{\rm pi}$ (for an external density
$n\,=\,1\,$cm$^{-3}$), orders of magnitude larger, and this length
scale gives the order of magnitude over which most electrons radiate
their energy through synchrotron radiation. An important question
therefore is whether Weibel turbulence, self-generated in the shock
precursor, can survive over the above time/length scales in order to
provide the turbulence in which synchrotron photons are produced, see
e.g.
\citet{1999ApJ...526..697M,1999ApJ...511..852G,2008ApJ...674..378C,2009ApJ...693L.127K,2013MNRAS.428..845L,2015JPlPh..81a4501L,2015MNRAS.453.3772L}
for discussions of this issue.

While additional instabilities or inverse cascade might potentially
help sustain or amplify this turbulence, one may also conceive a
simpler scenario, in which the power spectrum of magnetic fluctuations
at the shock contains long wavelength modes, with a long lifetime
$\propto k^{-3}$. As discussed in \citet{2013MNRAS.428..845L}, these
modes could be excited by the highest energy particles accelerated at
the shock front, which propagate over large distances in the
upstream. The gradual dissipation of the micro-turbulence by phase
mixing would then lead to a $\xi_B$ which now depends on some power of
the distance to the shock. The synchrotron spectrum produced by a
population of particles interacting with such a dissipative turbulence
bears some unique features, which could potentially be used as a
diagnostic of Weibel turbulence
\citep{2013MNRAS.428..845L,2015JPlPh..81a4501L}. A generic prediction,
in particular, is a significantly larger ratio of flux at high energy
to flux at low energy, because high energy photons are produced by
high energy electrons, which radiate on a shorter timescale, hence in
a stronger turbulence. Such a signature may have been detected in some
gamma-ray bursts, whose multi-wavelength light curve can be nicely
explained by a synchrotron spectrum resulting from a dissipative
turbulence with $\xi_B\,\propto\,(\omega_{\rm pi}x/c)^{-0.5}$.

%%%%%%%%%%%%%%%%%%%%%%%%%%%%%%%%%%%%%%%%%%%%%

\section{Macroscopic Simulations of Relativistic Shocks}\label{MC1} %111
The physics of relativistic collisionless shocks together with its
application to the phenomenology of powerful astrophysical sources,
are complex non-linear multi-scale problems, which have been addressed
with a variety of methods \citep[see
  e.g.][]{2011A&ARv..19...42B,2016RPPh...79d6901M}.  Significant
progress in our understanding of these phenomena has come through
particle-in-cell (PIC) simulations, which probe the full non-linear
relationship between the particles and the electromagnetic fields, see
\cite{SKL2015} and references therein.  Such simulations are however
computationally intensive and cannot yet explore the physics of shock
waves on the long macroscopic time and space scales on which the
source evolves, or on which radiation is produced. In this context,
the analytical theory that we have proposed in the previous section is
useful because it can be benchmarked against these PIC simulations on
small physical scales and extrapolated to the macrophysical scales of
interest. An alternative method is the numerical \mc\ technique which
we describe in this Section.

\mc\ methods offer a modeling technique that lies between PIC
simulations and \SA\ models.
In contrast to \SA\ descriptions which make a global diffusion
approximation, particles pitch-angle scatter locally with the
\mc\ technique \citep[e.g.,][]{EE84}.  The important difference is
that for the diffusion approximation to be valid, particle
distributions must be nearly isotropic in all frames and large flow
gradients must be avoided. While these approximations are good for
superthermal particles in \nonrel\ shocks, they are never good for
particles crossing \rel\ shock front. In the \mc\ code, particles
undergo a small \PAS\ event after a time small compared to a
gyroperiod. Thus the scattering is ``local" and highly anisotropic
distributions in sharp flow gradients can be treated directly. Of
course, the properties of the scattering are not calculated from first
principles, as done with PIC simulations. But scattering can be
directly parameterized, as so far done for \rel\ shocks
\citep[e.g.,][]{EWB2013}, or determined more fundamentally from
CR-induced plasma instabilities, \MFA\ (MFA), and an analytic
determination of scattering rates from the \SCly\ determined magnetic
turbulence power spectrum in \nonrel\ shocks
\citep[e.g.,][]{VBE2008,Bykov3inst2014}.

The analytical model described in the previous sections can be seen as
an effective two-fluid theory, describing how self-generated
electromagnetic turbulence affects and couples a fluid of superthermal
particles with a background plasma fluid.  As such, it can accurately
model the back-reaction of the accelerated particles and of the
background plasma on the turbulence.
In contrast, the \mc\ model is a one-component theory. The
\PAS\ approximation treats all particles the same and Fermi
accelerated particles arise smoothly from the background plasma which
is a \SCly\ described collection of particles rather than a separate
fluid.  The non-linear relationship between the Fermi accelerated
population and the bulk plasma is determined by iteratively solving
the energy-momentum tensor by forcing energy and momentum conservation
across the shock.

Although the relativistic \mc\ simulations that follow do not make a
distinction between the rest frame of the one-fluid plasma and the
rest frame of the turbulence, in contrast to Sec.~\ref{sec:weibel},
they will be generalized in the near future to a more realistic
setting in which the growth of turbulence and its rest frame are
dictated by microphysical considerations.  \mc\ simulations of
\nonrel\ shocks which incorporate these effects are described in the
following as an illustration.

Most \mc\ models thus far have assumed plane, steady-state shocks with
Fermi acceleration limited by free escape boundaries.  Parallel
\citep[e.g.,][]{EE84,EWB2016mfp}, oblique
\citep[e.g.,][]{1993ApJ...409..327B,SummerlinBaring2012} and
perpendicular \citep[e.g.,][]{2015ApJ...809...29T} shock geometries
have been modeled.

\newlistDE
The main features of the \mc\ technique are:
\listDE a large dynamic range can be modeled with reasonable computing
resources following particles consistently from thermal energies to
the highest CR energies with no constraint on the anisotropy of the
particle distribution;
\listDE the non-linear (NL) feedback on the shock structure from
efficient \Facc\ can be determined consistently with injection and the
production of ultra-high-energy CRs;
\listDE since particle anisotropies are treated consistently, no
distinction in the particle transport need be made between \nonrel,
\transrel, or \ultrarel\ shocks;
\listDE electrons and ions can be simultaneously injected and
accelerated  allowing effects depending on particle mass to be
determined \SCly; and
\listDE absolutely normalized radiation, from electrons and ions, can
be calculated as a straightforward function of model and environmental
parameters.

The main drawbacks of this method are that it is intrinsically steady
state, generally done for plain-parallel shocks, and thus far too
computationally slow to include in hydrodynamic evolutionary models
such as those for young SNRs using \SA\ descriptions of NL
\DSA\ \citep[e.g.,][]{EPSBG2007,LEN2012,ESPB2012,ZP2012}.

We describe here some physical results obtained with the
\mc\ technique which may help to understand the backreaction effects
of the accelerated particles on the shock flow at large scales which
can affect the particle spectra and radiation and which are difficult
to simulate yet with the PIC simulations.

\begin{figure}
\begin{center}
\includegraphics[width=0.8\columnwidth]{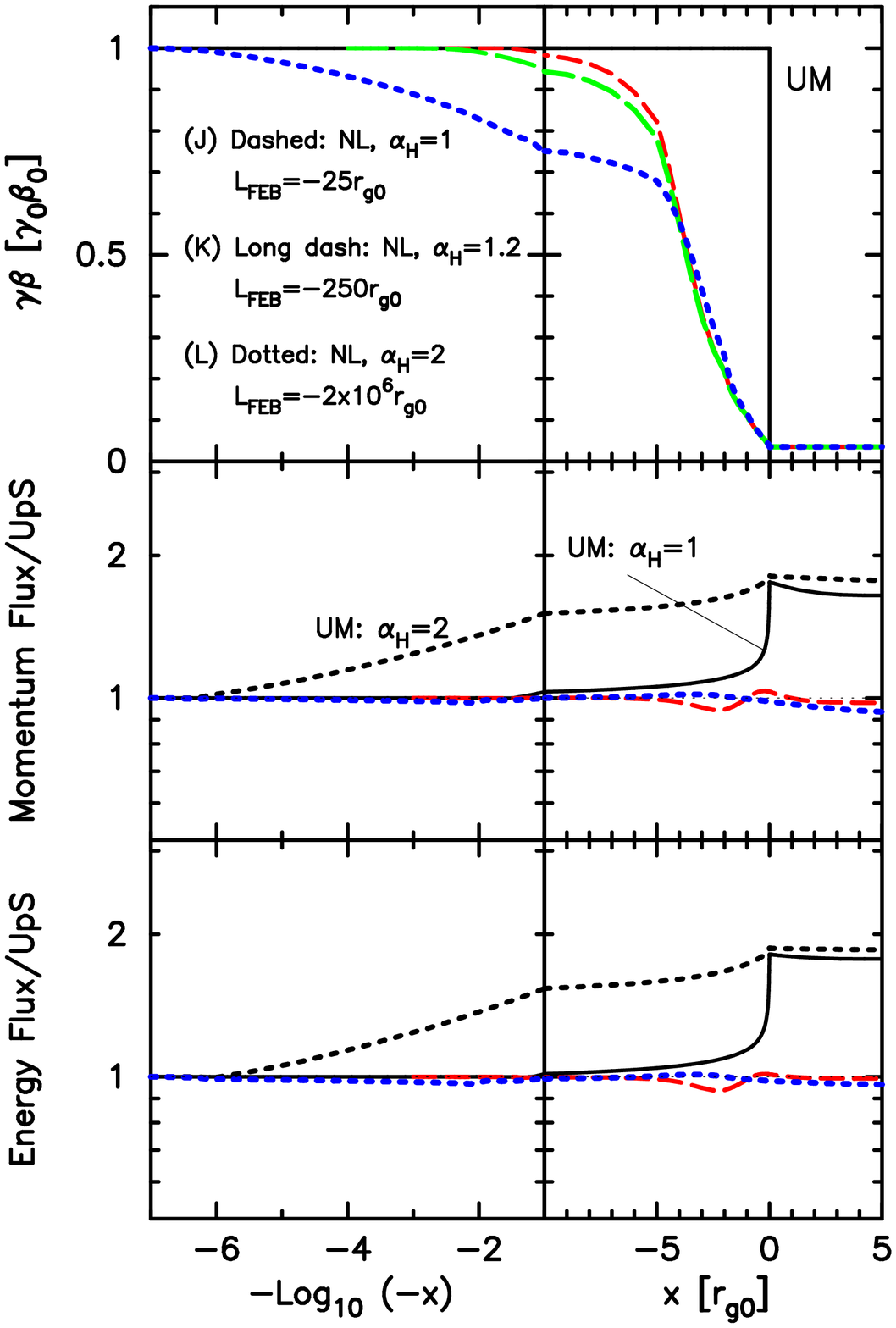}  % fig 5
\caption{The top panels show the non-linear (NL) shock structure in
  terms of the inverse density, $\GBeta/(\GBetaZ)$, for shocks with
  mean free paths having different momentum dependence illustrated in
  Fig.~\ref{fig:fig8_mfp}. This figure is adapted from figure~7 in
  \citet{EWB2016mfp} and that paper has full details. In all cases,
  the shock \Lor\ factor is $\gamZ=10$ and curves labeled UM are for
  unmodified shocks where the shock structure is
  discontinuous.}\label{fig:prof3}
\end{center}
\end{figure}

\subsection{PIC vs Monte Carlo simulations}
At present, the self-consistent problem of relativistic collisionless
shock structure, turbulence production, and particle acceleration can
only be solved with full scale PIC simulations.
A very important result of these simulations is the demonstration that
the Weibel instability, as well as current-driven filamentation
instabilities, produce short-wavelength turbulence which can scatter
accelerated particles \citep[see
  e.g.][]{Spitkovsky2008,2011A&A...532A..68P,SSA2013,2014MNRAS.440.1365L}.
In the simulations performed thus far, the self-generated turbulence
is seen to fill a layer of a few hundred ion skin depths close to the
subshock.
Another important result is that PIC simulations have shown that,
regardless of the background magnetic field geometry, unmagnetized
relativistic shocks can be efficient particle accelerators converting
$\gsim 10$\% of their ram pressure into accelerated particles
\citep[e.g.][]{SSA2013}.

Despite these successes, the limitations on box size, run time, and
dimensionality \citep[see][]{1998ApJ...509..238J,VBE2008} inherent in
PIC simulations force the consideration of less fundamental
descriptions. This is especially true for \nonrel\ and
\transrel\ shocks where the precursors may extend over the scales of
hundreds or thousands of gyroradii of maximal energy CRs.
By parameterizing the particle transport, the \mc\ technique we
discuss can simulate the non-linear shock structure on astrophysically
significant scales for arbitrary shock speeds
\citep[see][]{EBJ95,SummerlinBaring2012,EWB2016mfp}.
In the case of non-relativistic shocks, the \mc\ model can
simultaneously include the non-linear shock structure, accelerated
particle production, and magnetic turbulence generation accounting for
both short scale and long-wavelength instabilities produced by the
anisotropic CR particle distributions
\citep[e.g.][]{VBE2009,Bykov3inst2014}.
It is even possible to include super-diffusive propagation
\citep[i.e. the so-called \levy-walk models, e.g.][]{ZP2013}
consistently in the shock precursor \citep[][]{BEO2017}.

In principle, microphysical parameters determined with PIC simulations
can be implemented in the \mc\ code to allow the analysis of
large-scale CR modified shocks. While this will allow for a large
increase in dynamic range, it remains uncertain whether or not
self-generated CR-driven instabilities occur on large scales,
particularly for fully \rel\ shocks \citep[see
  e.g.][]{SagiNakar2012,2014MNRAS.440.1365L}.
In the next section we present some \mc\ results showing the
large-scale macroscopic structure of relativistic shocks. We also
include a discussion of \transrel\ and \nonrel\ shocks to emphasize
that the differences between shocks of varying speed is mainly one of
analysis; mathematical techniques separate shocks more than the
underlying physics which must be continuous through the
\transrel\ range.

\section{Non-linear Shock Structure}\label{NLMC1}
It is reasonably certain from direct observations that \nonrel\
shocks can inject and accelerate ions efficiently, while the
electron acceleration efficiency, at least in interplanetary shocks,
is still uncertain \citep[e.g.,][]{2016A&A...588A..17D}. While no
direct observations exist for \rel\ shocks, many models,
particularly those for \GRBs\ (GRBs), assume \Facc\ is efficient
with $\gsim 10\%$ of the shock ram kinetic energy placed in
nonthermal particles. In many cases, this energy is assumed to be
given to electrons since they produce readily observable \syn\ and
\ICd\ emission.
To our knowledge, there are no \SC\ models of \FoFSA\ that place
more energy in electrons than ions. Unless shocks are accelerating a
pair dominated plasma  whatever energy fraction is assumed for
electrons will be equaled or exceeded by ions even if they do not
produce observable radiation.

If \Facc\ is efficient, the shock structure must be modified by the
back pressure of CRs as they scatter in the shock precursor. This is
illustrated in Fig.~\ref{fig:prof3} which is adapted from figure~7
in \citet*{EWB2016mfp}.
The shocks shown in Fig.~\ref{fig:prof3} are \rel\ with a \Lor\
factor $\gamZ=10$. For a given acceleration efficiency, \nonrel\
shocks will have a more pronounced precursor structure because it is
easier for CRs to scatter upstream in a \nonrel\ flow.
The \mc\ code determines the momentum and energy conserving shock
structure by iteration, as described in \citet{EBJ96,EWB2013}.

\begin{figure}
\includegraphics[width=0.8\columnwidth]{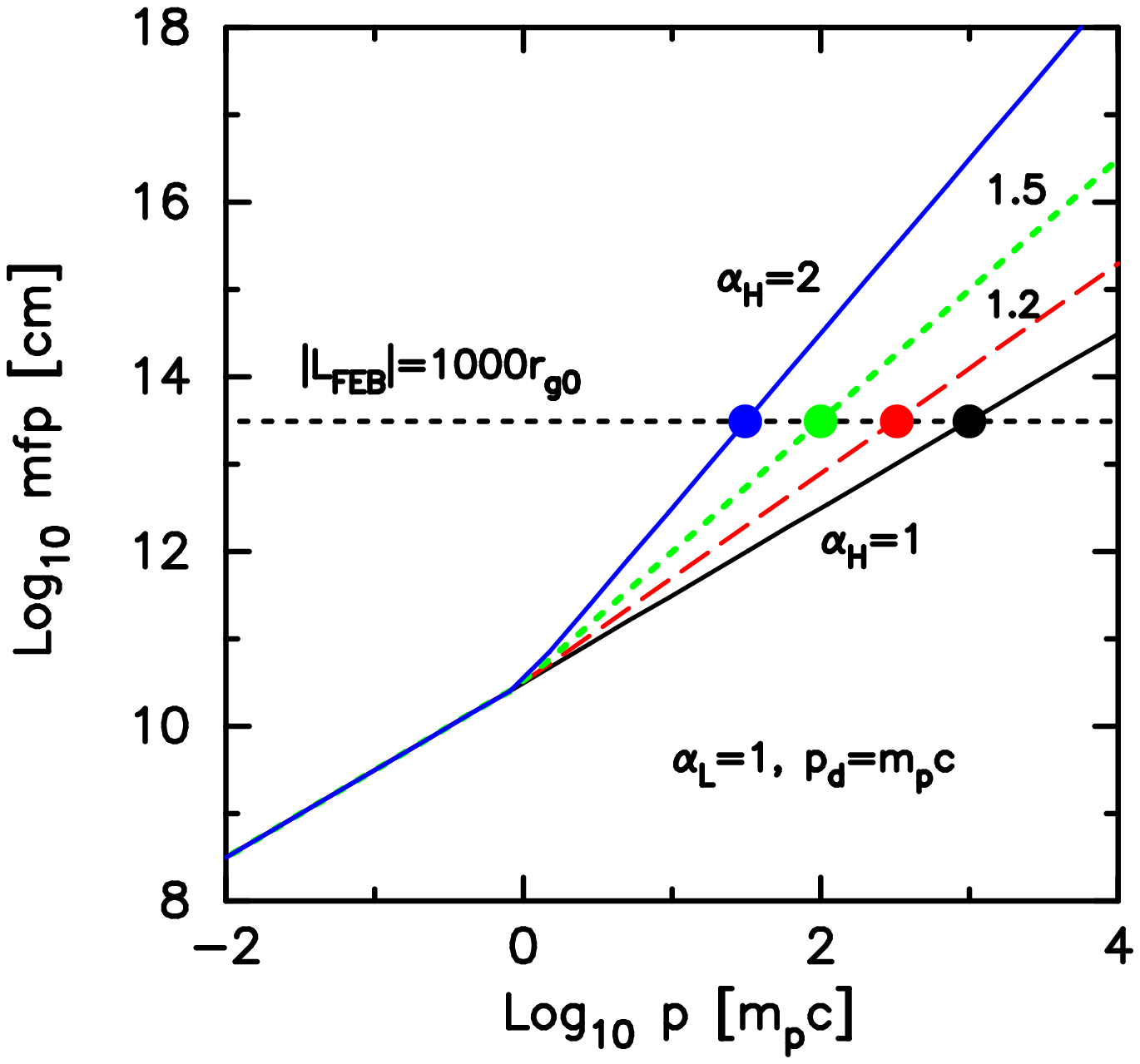}
\caption{Mean free paths for the \SC\ shocks shown in
  Fig.~\ref{fig:prof3} \citep[see][]{EWB2016mfp}. The positions of the
  free escape boundaries (FEBs) are shown and it is noted that the
  momenta at which $\Lmfp(\pFEB) \simeq |\Lfeb|$ (solid dots) are such
  that $\pFEB/\pmax$ is a strong increasing function of $\alfH$.
\label{fig:fig8_mfp}}
\end{figure}

\begin{figure}
\includegraphics[width=0.8\columnwidth]{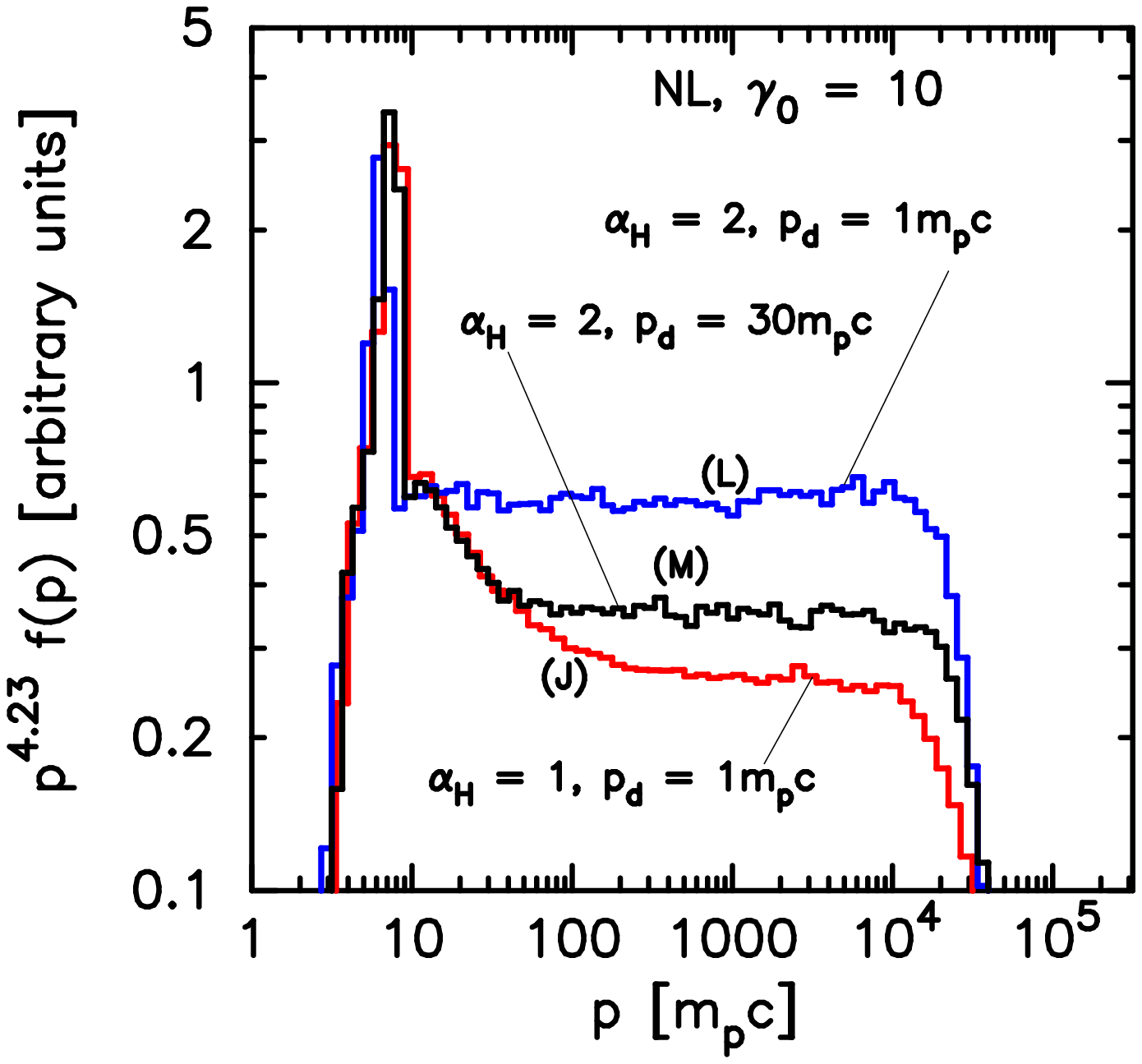}
\caption{Phase-space distributions with various indexes $\alfH$ and
  $p_d$, as indicated \citep[see][]{EWB2016mfp}.  Note the effect
  $p_d$ has on the \Facc\ efficiency in the top two curves both with
  $\alfH=2$.  Model (J) is simulated with $\alfH=1$ and the upstream
  FEB is at $-25\,\rgZ$. To have essentially the same $\pmax$, model
  (L) with $\alfH=2$ has a FEB at $-2\xx{6}\,\rgZ$. All spectra are
  calculated at $x=0$ in the rest frame of a shock with $\gamZ$ = 10.
\label{fig:fig11_mfp}}
\end{figure}

\subsection{Iteration of Shock Structure}
If we assume that the shock is parallel and if we ignore the effects
of escaping momentum and energy fluxes, the shock-frame, steady-state
flux conservation relations are: %%
\begin{equation} \label{eq:NumFlux}
\Fnum(x) = \gamma(x) n(x) \beta(x)  = \FnumZ \ ,
\end{equation}
\begin{eqnarray} \label{eq:PxFlux}
\Fpx(x) &=& \gamma^2(x) \beta_x^2(x) [e(x) + \Pxx(x)] + \Pxx(x) =
\FpxZ,
\nonumber \\
\end{eqnarray}
and
\begin{equation} \label{eq:EnFlux}
\Fen(x) = \gamma^2(x) \beta_x(x) [e(x) + \Pxx(x)] = \FenZ \ .
\end{equation}
Here, $\FnumZ$, $\FpxZ$, and $\FenZ$ are the far upstream number,
momentum, and energy fluxes, respectively, and $\Pxx$ is the
$xx$-component of the pressure tensor, $\mathcal{P}$.
Given the input quatities, $\FnumZ$, $\FpxZ$, and $\FenZ$, these
three equations have four unknowns: $n(x)$, $\gamma(x)$, $e(x)$, and
$\Pxx(x)$, so an additional constraint is needed for solution. We
use the adiabatic and gyrotropic equation of state:
\begin{equation}\label{eq:EqState}
e(x) = \frac{P(x)}{\Gamma(x) -1} + \rho(x) c^2 \ ,
\end{equation}
where $e(x)$ is the total energy density, $\rho(x) c(x)^2$ is the rest
mass energy density, $\Gamma(x)$ is the adiabatic index, and
$P(x)=\Pxx$ is the isotropic scalar pressure, i.e., $P =
Tr(\mathcal{P})/3$ \citep[see][for more details]{DoubleEtal2004}.  All
of these quantities, along with the bulk flow \Lor\ factor (also
called the shock profile) $\gamma(x)$, vary with position $x$ in the
precursor.

The solution is accomplished in the \mc\  code by iteration.
Starting with a shock profile, i.e., $\gamma(x)$, the simulation is
run and $\Fnum(x)$, $\Fpx(x)$, and $\Fen(x)$ are measured at every
$x$. If the measured fluxes don't match the known values $\FnumZ$,
$\FpxZ$, and $\FenZ$ to within some error limits, $P(x)$, which is a
local frame quantity, is determined using Eq.~(\ref{eq:PxFlux}) or
(\ref{eq:EnFlux}). Using this new $P(x)$,
Eqs.~(\ref{eq:NumFlux})--(\ref{eq:EnFlux}) are used to predict the
next shock profile, i.e., the next $\gamma(x)$. This process is
iterated until the fluxes are conserved. An example from data
presented in \citet{EWB2016mfp} is shown in Fig.~\ref{fig:prof3}.
In contrast with \SA\ techniques where multiple solutions can be
obtained \citep[see][for a review]{Drury83}, unique solutions always
result with the \mc\ approach because of the non-linear (NL) feedback
between the thermal injection and the shock structure.

The above procedure becomes significantly more complicated with an
oblique magnetic field. In this case, there are  $x$-dependent
parallel and perpendicular components of the magnetic field and
equations (\ref{eq:NumFlux})---(\ref{eq:EnFlux}), are replaced by
six equations which include magnetic pressure. This system has been
derived by \citet{DoubleEtal2004}.
\SCc\ solutions have been found for NL oblique \nonrel\ shocks
\citep[e.g.,][]{EBJ96}, oblique \rel\ unmodified shocks
\citep[e.g.,][]{ED2004}, but we know of no solutions, other than
from PIC simulations, for NL oblique \rel\ shocks.

\subsection{Particle Transport and Magnetic Turbulence}
Pitch-angle particle scattering and convection are determined in one
of two ways in the \mc\ code. In the simplest scheme the \mfp\
$\lambda(x,p)$ is a fully parameterized function of particle
momentum $p$ and position $x$ relative to the subshock at $x=0$.
This method has been used extensively in previous \nonrel\
applications \citep[e.g.,][]{BE99} and is currently used for \rel\
shocks.

In Fig.~\ref{fig:fig8_mfp} we show examples of the scattering mfp
for \rel\ shocks ($\gamZ=10$) where the momentum dependence has been
varied as follows \citep[see][for full details]{EWB2016mfp}:
\begin{eqnarray} \label{eq:mfp}
\Lmfp(p) &=& \etamfp \cdot \rg(\pD) \cdot (p/\pD)^{\alfL} \quad
\mathrm{for} \quad p < \pD
\nonumber \\
&=& \etamfp \cdot \rg(\pD) \cdot (p/\pD)^{\alfH} \quad \mathrm{for}
\quad p \ge \pD \ .
\end{eqnarray}
Here $p$ is the particle momentum in the local frame, $\alfL \le 1$,
$\alfH \ge 1$, $\pD$ is a dividing momentum between the low- and
high-momentum ranges, $\rg(\pD)=\pD c/(eB_0)$ is the gyroradius for
a proton with local-frame momentum $\pD$ in the background field
$B_0$, and $\etamfp \ge 1$ is a parameter that determines the
strength of scattering.
More complicated forms for $\Lmfp(p)$ can be used in the \mc\
simulation to model \SA\ and/or PIC results where available. Note
that $\Lmfp(p)$ was taken to be position independent in  the results
shown in Figs.~\ref{fig:fig8_mfp} and \ref{fig:fig11_mfp}.

The Bohm limit is described by $\etamfp = \alfL = \alfH = 1$, and
the conditions $\alfL \le 1$ and $\alfH \ge 1$ ensure that $\Lmfp
\ge \etamfp \rg$ for all $p$.\footnote{The case with
$\etamfp=\alfL=\alfH=1$ is often referred to as ``Bohm diffusion"
however particle trajectories are calculated in the \mc\ code
without making a diffusion approximation.}
In the simple geometry of a plane-parallel, steady-state shock,
results scale with $\etamfp$. Large values of $\etamfp$ imply weak
scattering with long length and time scales.

In Fig.\ref{fig:fig8_mfp}, $\alfH=2$ is used as a typical value
predicted by the analytical theory of Sec.~\ref{sec:weibel} of
particle transport in small-scale micro-turbulence.

Regardless of the form for $\Lmfp$, pitch-angle scattering is
modeled as follows: after a time $\deltime \ll \gyrotime$
($\gyrotime$ is the gyro-period) a particle scatters isotropically
and elastically in the local plasma frame through a small angle
$\delAng \le \delMax$.
The maximum scattering angle in any scattering event is given by
\citep[see][]{EJR90}
\begin{equation} \label{eq:Tmax}
\delMax = \sqrt{6 \deltime/t_c} \ ,
\end{equation}
where $\deltime=\gyrotime/N_g$, $t_c=\Lmfp/v$ is the collision time,
$v$ is the particle speed in the local frame, and $N_g$ is a free
parameter. Typically, $N_g$ is chosen large enough to produce
fine-scattering results that do not change substantially as $N_g$ is
increased further \citep[see][for examples with values of $N_g$
producing large-angle scattering]{SummerlinBaring2012}.

All particles are injected far upstream with a thermal distribution
and scatter  and convect into and across the subshock into the
downstream region. A critical approximation used to date in the \mc\
simulation is that the subshock is transparent, i.e., no attempt is
made to describe the effects of a cross-shock potential, amplified
magnetic turbulence, or other effects that may occur in the viscous
subshock layer.
This approximation has been relaxed in recent work with the
parameterization of energy transfer between ions and electrons at
the subshock and an injection threshold
\citep[][]{WarrenEllison2015}.

Upon interacting with the downstream plasma, some particles will
gain energy sufficient to allow them to scatter back across the
subshock and be further accelerated. The fraction of particles that
are injected, i.e., do manage to re-cross the subshock is determined
stochastically and constitutes a ``thermal leakage injection" model
once the  further assumption is made that the subshock is
transparent.
This injection model requires no additional parameters or
assumptions once the scattering mean free path is defined in
equation~(\ref{eq:mfp}).

The additional parameter implemented by \citet{WarrenEllison2015} to
model radiation spectra from unmagnetized relativistic shocks is
$\Fion$ which is the fraction of far upstream ion ram kinetic energy
transferred to electrons. The far upstream kinetic energy of an ion
is
$(\gamZ -1) m_i c^2$, where $m_i$ is the ion mass. When an ion
crosses the subshock from upstream to downstream for the first time,
a part of this energy equal to $\Fion (\gamZ -1) m_i c^2$ is removed
from it.  The total energy removed from all of the ions is divided
equally among electrons and added to their energy as they cross the
subshock into the downstream region for the first time. This
parameterisation was implemented to account for the effect of the
energy exchange between the ions and electrons in the subshock which
is clearly seen in Fig.~11 from \citet{SSA2013} where PIC
simulations of a low magnetization ($\sigma < 10^{-5}$) ion-electron
plasma shock were presented.

\subsection{Particle Escape}
While the majority of shocked  particles convect downstream and
don't  re-cross the subshock to experience \Facc, a small fraction
will continue to be accelerated  until they run out of acceleration
time or their diffusion length becomes comparable to the shock size.
Both the acceleration rate and the effective shock size depend
critically on self-generated turbulence which will always have a
finite extent.
The finite extent of self-generated turbulence guarantees that
particle escape will play a role in \Facc.

In quasi-steady-state shocks, such as the Earth bow shock, or
long-lived shocks, such as SNR blast waves,  the effective shock
size limits acceleration (except at earliest times for SNRs) and the
escape of particles from the shock becomes an important part of the
Fermi process. In fact, in high Mach number \nonrel\ shocks, it
isn't possible to conserve energy unless escape occurs and this has
been modeled with a free escape boundary (FEB). The steeper spectrum
produced by \Facc\ in \rel\ shocks obviates the energy conservation
argument, but the finite extent of self-generated turbulence means
particle escape will occur here as well.

Because self-generated turbulence will decay in the downstream
region, particle escape can occur from downstream as well as
upstream from the precursor.\footnote{In the often used
approximation of an infinite, plane shock, particles do not
``escape" downstream unless a boundary is imposed. Particles convect
away from the shock but always have a finite, but increasingly
small, probability of scattering back to the subshock.}
The \mc\ simulation limits acceleration with an upstream and/or
downstream FEB, although lateral escape can also be modeled. Lateral
escape may be important in objects as diverse as the Earth bow shock
and GRB jets.
How escape actually occurs will depend on the detailed geometry of
the shock but all these scenarios produce a corresponding $\pmax$
with no differences large enough to be discerned from current
observations.
Even though the \mc\ model is steady-state, the acceleration time of
individual particles is calculated and \Facc\ can also be limited by
setting a maximum time \citep[see][for a discussions of particle
escape]{Achterberg2001,Drury2011}.

An upstream FEB has been used extensively in models of CR production
in young SNRs \citep[e.g.,][]{EPSBG2007,MAB2009,CAB2010} where there
is typically a SNR age where a transition occurs between $\pmax$
being determined by the remnant age to being determined by the
finite shock radius. We note that there is direct observational
evidence for upstream escape at the quasi-parallel Earth bow shock
\citep[e.g.,][]{Trattner2013}.

Using Eq.~(\ref{eq:mfp}) to vary the momentum dependence of
$\Lmfp(p)$ we show some examples in Fig.~\ref{fig:fig11_mfp}. These
are shock-frame, phase-space distributions for \rel\ shocks with
$\gamZ=10$. For these examples, the size of the precursor (i.e., the
position of the upstream FEB) has been varied to produce
approximately the same $\pmax$.
The model parameters are described in detail in \citet{EWB2016mfp}
but it is sufficient to say that the precursor size depends strongly
on the momentum dependence of $\Lmfp(p)$. For Model (J) with
$\alfH=1$, the upstream FEB is at  $-25\,\rgZ$, while Model (L) with
$\alfH=2$ and  essentially the same $\pmax$,  has a FEB at
$-2\xx{6}\,\rgZ$. Here $\rgZ=m_p u_0 c/(e B_0)$ where $u_0$ is the
shock speed, $B_0$ is the far upstream magnetic field, and $m_p$,
$c$, and $e$ have their using meanings.
A value of $\alfH>1$ implies weak turbulence but this also means
that energetic particles will be able to scatter well into the
precursor opening the possibility of self generation of precursor
turbulence.

\subsection{Thermal Particle Injection}
A collisionless shock moving through ambient material will heat all
of the material and inject some fraction of that material into the
\Facc\ mechanism. While it is clear that a background superthermal
population will always be accelerated \citep[e.g.,][]{BO78}, there
is compelling evidence that some fraction of the cold unshocked gas,
upon being heated, will be Fermi accelerated as well
\citep[e.g.,][]{BE87,EMP90,SSA2013}. We characterize any process
that turns cold, unshocked material into superthermal particles,
thermal particle injection.

\newlistDE

Thermal particle injection (TPI) has been considered a particularly
difficult part of \Facc\ from both observational and theoretical
aspects.
By definition, thermal particles are always highly anisotropic at
shocks of any speed, making injection difficult to describe with
\SA\ models based on the diffusion-advection equation.
Injection can be treated more directly with the \mc\ code since
there are no constraints on anisotropy.
This makes it possible to describe injection in an internally \SC\
way within the assumptions made to describe particle transport with
no additional assumptions.

However, as seen in PIC simulations, the subshock layer will be highly
turbulent \citep[e.g.,][]{SSA2013,CPS2015} making the injection
process fundamentally complex. While the simple approximations for
particle scattering made by the \mc\ model cannot describe the full
complexity of TPI, \mc\ results have been shown to be reasonably
consistent with PIC simulations of \rel\ shocks.
Using similar parameters, the \mc\ simulation can reasonably
reproduce the NL shock structure, particle spectral shape, and
overall acceleration efficiency obtained by the PIC simulation
\citep[see][for details]{EWB2013}. We know  of no semi-analytic
method that can do this  for \rel\ shocks.  In effect, the \mc\
injection approximations average over the complex microscopic
details while ensuring that global momentum and energy conservation
is maintained.
These approximations allow parameterizations of PIC results that can
then be applied on larger CR momentum, length, and time scales than
are currently possible with PIC simulations.

\begin{figure} % fig 7  fig8 from \citet{WarrenEllison2015}
  \includegraphics[width=0.8\columnwidth]{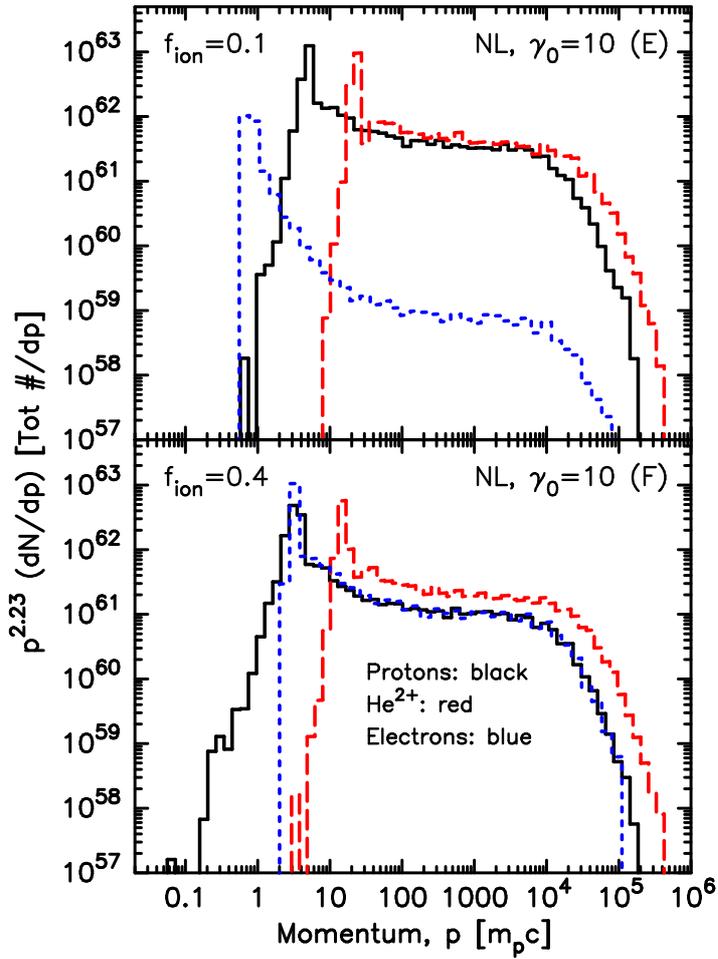}
  \caption{Non-linear downstream spectra for models of relativistic
    ion-electron shocks of $\gamZ$ = 10 with $\Fion=0.1$ and
    $\Fion=0.4$ simulated by \citet{WarrenEllison2015}.  In both
    panels the solid (black) curves are protons, the dashed (red)
    curves are \HeT, and the dotted (blue) curves are electrons.
\label{fig:dNdp_fion}}
\end{figure}

\begin{figure} % fig 9
\includegraphics[width=0.8\columnwidth]{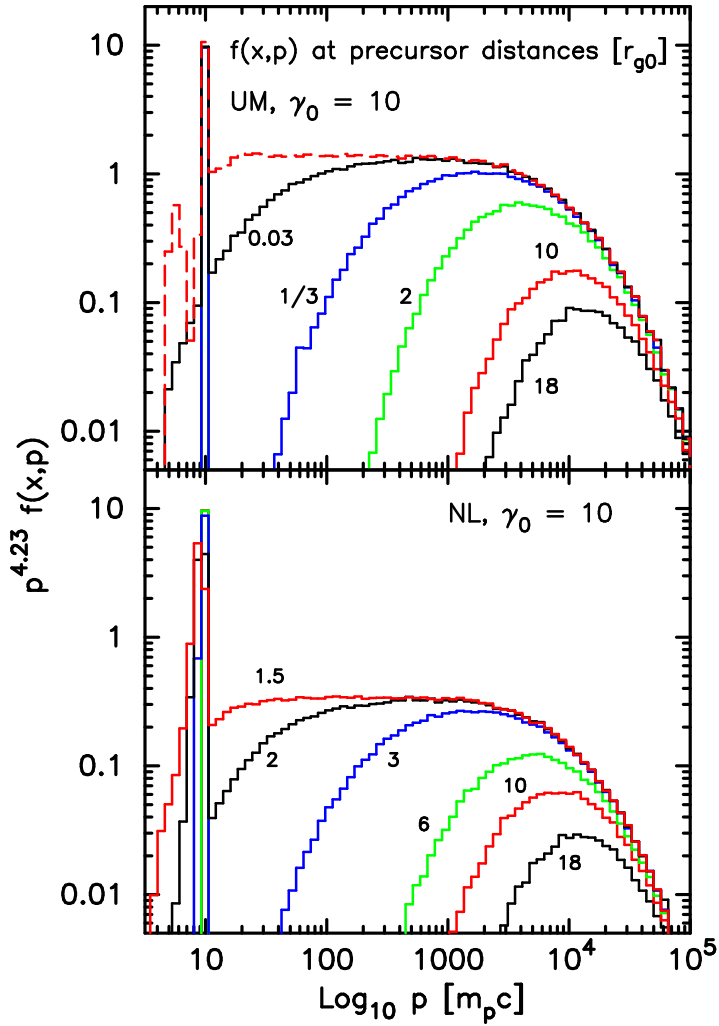}
\caption{The distribution function simulated with the \mc\ technique
  [represented as $p^{4.23}f(x,p)$] at the different distances in the
  shock precursor. The distance (which is measured in $\rgZ$) is
  indicated by a label on the corresponding curve. The simulation
  results for an unmodified shock (UM) are shown in the top panel
  while $p^{4.23}f(x,p)$ for a simulated non-linear shock (NL) where
  momentum and energy are conserved is shown at the bottom
  panel.\label{fig:pen_fp}}
\end{figure}

\begin{figure} % fig 9
\includegraphics[width=0.8\columnwidth]{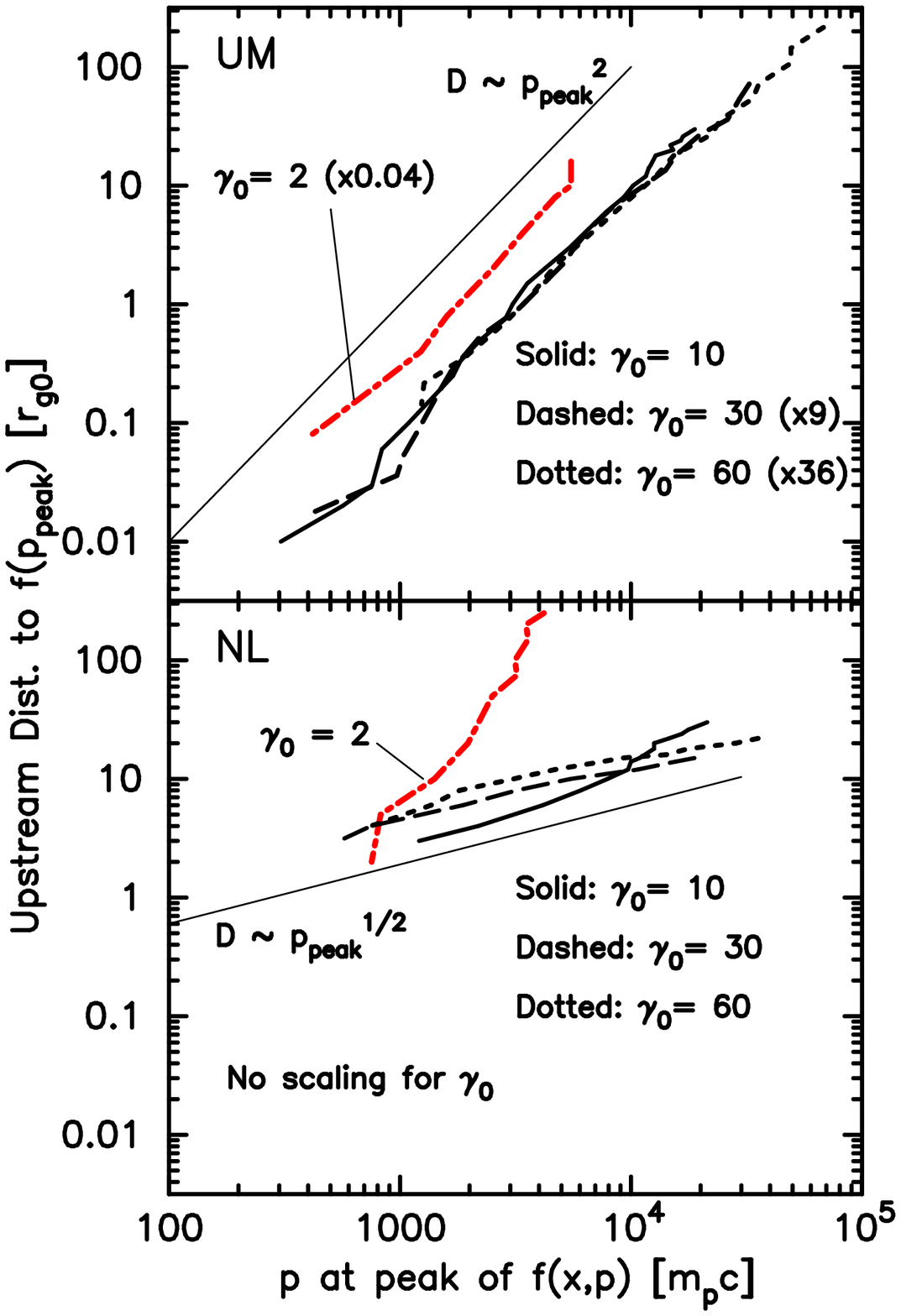}
\caption{The characteristic precursor scale derived as the distance
  $\Dpen(\Ppeak)$ from the shock to the precursor position where
  $p^{4.23}f(x,p)$ peaks.  The momentum $\Ppeak$ corresponds to the
  peak of $p^{4.23}f(x,p)$ in Fig.~\ref{fig:pen_fp}.  We plot $\Dpen$
  vs. $\Ppeak$ for shocks with varying \Lor\ factors.  In the top
  panel the results for unmodified shocks (UM) are presented, while
  the bottom panel shows the results for non-linear shocks (NL) see
  Fig.~\ref{fig:pen_fp}.
  \label{fig:pen_sc}}
\end{figure}

\subsection{Precursor Structure}
In order for \Facc\ to work, some downstream particles must scatter
back across the shock into the upstream region creating a precursor.
While the length of this precursor is a strong inverse function of
shock \Lor\ factor, it must be present on some scale if CRs are
accelerated beyond the initial boost they obtain in their first
interaction with the shock.

We characterize the precursor scale with a penetration depth,
$\Dpen(\Ppeak)$, where $\Ppeak$ is the momentum where $p^{4.23} f(p)$
peaks. Examples are shown in Fig.~\ref{fig:pen_fp} for an unmodified
(UM) shock (top panel) and the NL shock (bottom panel) where momentum
and energy are conserved and the shock is smoothed. While it is clear
that the momentum of the peak, $\Ppeak$, is not precise---the
precursor distributions are not delta functions---there is a clear
difference in $\Dpen(\Ppeak)$ between the two cases.

In Fig.~\ref{fig:pen_sc} we plot $\Dpen$ vs. $\Ppeak$ for shocks with
varying \Lor\ factors. In the UM shocks (top panel), $\Dpen \propto
\Ppeak^2$ in all cases. Note that the curves in the top panel are
scaled to the $\gamZ=10$ result with a normalization factor
$(\gamZ/10)^2$ as shown. For a given $p$, the penetration depth scales
as $1/\gamZ^2$, as expected for an UM flow. The NL shocks show
something very different. For $\gamZ=10$, 30, and 60, $\Dpen \propto
\Ppeak^{1/2}$. It is much easier for low momentum particles to diffuse
upstream once the precursor develops. At the highest $p$ shown in
Fig.~\ref{fig:pen_sc}, the penetration depth becomes similar for the
UM and NL cases.
Note that the \transrel\ UM $\gamZ=2$ shock obeys the $\Dpen \propto
\Ppeak^2$ scaling but doesn't conform to the $1/\gamZ^2$ scaling.

\Facc\ in \rel\ shocks has been suggested as a possible source of
ultra-high-energy CRs \citep[e.g.,][]{KW2005}, and \rel\ shocks may be
responsible for non-thermal radiation observed from a number of
astrophysical sources including \GRBs\ (GRBs), type Ibc supernovae,
and extra-galactic radio jets. In order to explain this CR production
and radiation, electron and ion acceleration to \ultrarel\ energies
must be modeled.
\PIC\ simulations concentrate on microphysical aspects of \rel\
shocks but lack the dynamic range needed to model \NL\ shock
structure and \Facc\ from thermal injection energies to CR energies
above $\sim 10^{15}$\,eV.
On the other hand, semi-analytic descriptions of magnetic field
amplification and particle scattering can cover a large dynamic range
but have not yet included non-linear effects from efficient Fermi
acceleration. We believe the \mc\ technique offers a way to combine
insights gained from \SA\ studies with PIC results, such as the
transfer of ion energy to electrons, in a NL, broadband model capable
of producing radiation that can be compared with observations
\citep[e.g.,][]{WEBL2016,WEBN2016}.

\subsection{Lessons from non-relativistic shock models:
Scattering Center Speed}
Thus far, the calculation of self-generated turbulence in NL
\rel\ shocks has only been done with PIC
simulations. \SAn\ calculations have been performed in unmodified
\rel\ shocks, as discussed earlier.  As a prelude to future
relativistic shock \mc\ calculations, we show here some examples for
NL \nonrel\ shocks performed recently by \citet{Bykov3inst2014}.
Despite the physical differences between \rel\ and \nonrel\ shocks
discussed above, the \nonrel\ results can serve as a useful guide.

\begin{figure}
\includegraphics[width=0.8\columnwidth]{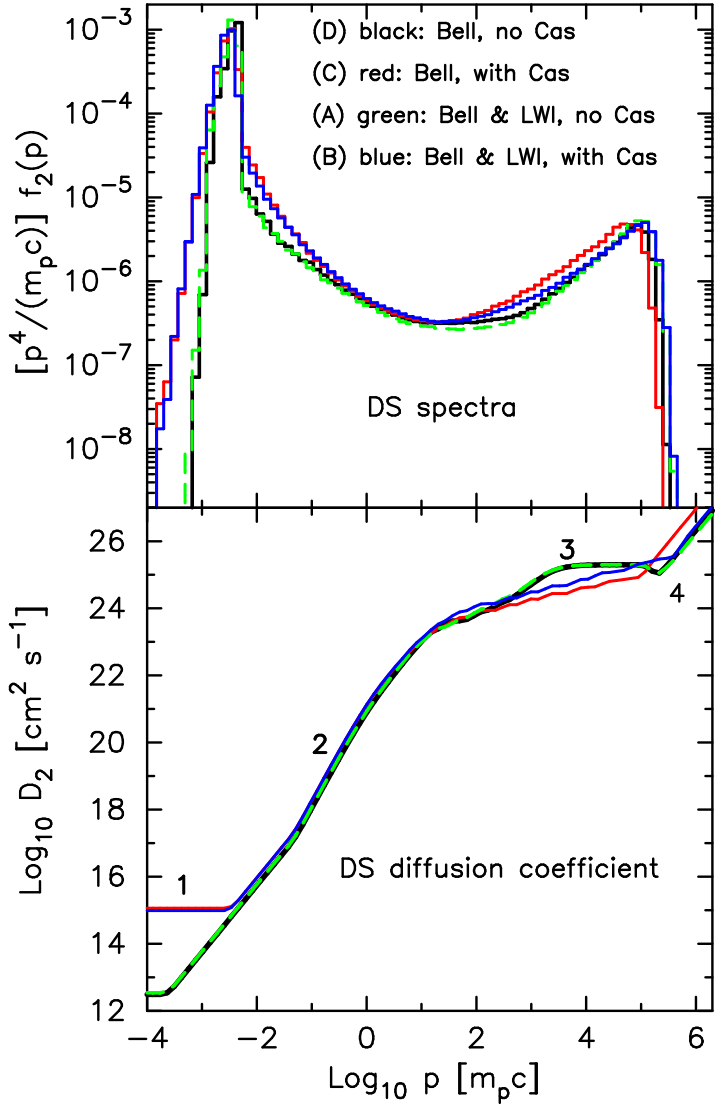}
\caption{Phase-space particle distributions, $[p^4/(m_pc)]f_2(p)$ (top
  panel), and diffusion coefficients (bottom panel) calculated
  downstream from the subshock in the \NL\ \mc\ model for shock
  velocity 5,000\,\kmps, the far upstream density $n_{0}=0.3\,
  \mathrm{cm}^{-3}$, and magnetic field $B_0$ = 3 $\mu$G. The particle
  spectra and diffusion coefficients calculated consistently for
  different models of magnetic turbulence
  \citep[from][]{Bykov3inst2014}.} \label{fig:3InstD}
\end{figure}

The NL, \nonrel\ \mc\ model includes magnetic field amplification
from resonant instabilities, as well as short-wavelength
\citep[e.g.,][]{Bell2004,PLM2006,ab09} and long-wavelength
\citep[e.g.,][]{BER2012,SchureEtal2012} non-resonant instabilities.
In Fig.~\ref{fig:3InstD}, taken from \citet{Bykov3inst2014}, we show
downstream particle spectra (top panel) for examples with and
without the long-wavelength, non-resonant instability. The examples
shown were simulated  with and without the turbulence cascade, as
indicated.
It is important to note that in these \nonrel\ results, the scattering
mean free path is \SCly\ determined and has an arbitrary $p$ and $x$
dependence. The \SCly\ determined diffusion coefficient is shown in
the bottom panel in Fig.~\ref{fig:3InstD}, again taken from
\citet{Bykov3inst2014}. It is clear from Fig.~\ref{fig:3InstD} that
the mfp does not have a simple Bohm-like momentum dependence.
A very important distinctive feature of these \nonrel\ results is
the spectral hardness at high $p$.  Such hard spectra are not
expected in  \rel\ shocks, as shown in Fig.~\ref{fig:dNdp_fion}.
The hard CR spectrum in NL, \nonrel\ shocks provides  the free
energy needed for strong amplification of magnetic turbulence in the
shock precursor. It remains to be seen how turbulence will generated
in \rel\ shocks.

Self-generated magnetic turbulence is essential for \Facc\ to work
since particles must scatter nearly elastically off this turbulence
to cross the subshock several times and be accelerated above thermal
energies.
An important aspect of this is that the turbulence may move through
the bulk plasma so the effective compression ratio for \Facc,
$\Reff(x) = \Vscat(x)/u_2$, may differ from the compression ratio of
the bulk plasma, $R = u(x)/u_2$, where $u_2$ is the downstream flow
speed in the shock rest frame.

\begin{figure}
\includegraphics[width=0.8\columnwidth]{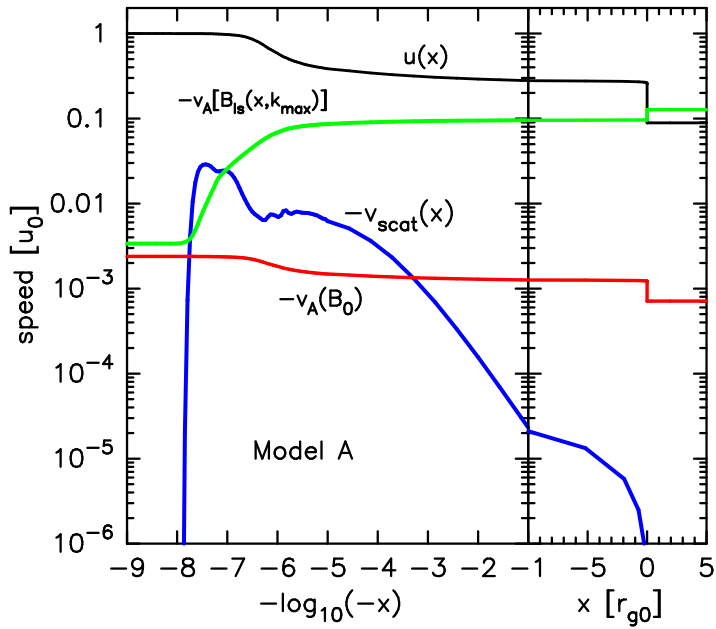}
\caption{The scattering center velocity profile (solid line) derived
  from the \NL\ \mc\ model for shock velocity 5,000\,\kmps, the far
  upstream density $n_{0}=0.3\, \mathrm{cm}^{-3}$, and magnetic field
  $B_0$ = 3 $\mu$G. We also show the flow velocity profile (dashed
  line) and \alf\ velocities calculated with the amplified magnetic
  field $\Bls(x)$ (dotted line) and the initial field $B_0$
  (dot-dashed line) \citep[from][]{Bykov3inst2014}.} \label{fig:ScatV}
\end{figure}

Fig.~\ref{fig:ScatV} shows an example from a \nonrel\ shock
calculation where $\Vscat$ has been  determined \SCly\ with the \mc\
simulation. While most \SA\ calculations assume $\Vscat$ is the
\alf\ speed, $\Valf(x)$, calculated either with the ambient magnetic
field or the amplified field, the \SC\ results obtained in
\citet{Bykov3inst2014} show that it can differ substantially from
$\Valf(x)$ regardless of the magnetic field assumed.

The velocity of the scattering centers, i.e. the Weibel frame, has
been determined analytically in the relativistic regime in
Sec.~\ref{sec:weibel}. This motion is of course implicitly included
\SCly\ in PIC simulations, and it will be incorporated in future
\mc\ simulations of relativistic shocks.

\begin{figure} % fig 8  fig9 from \citet{WarrenEllison2015}
\includegraphics[width=0.8\columnwidth]{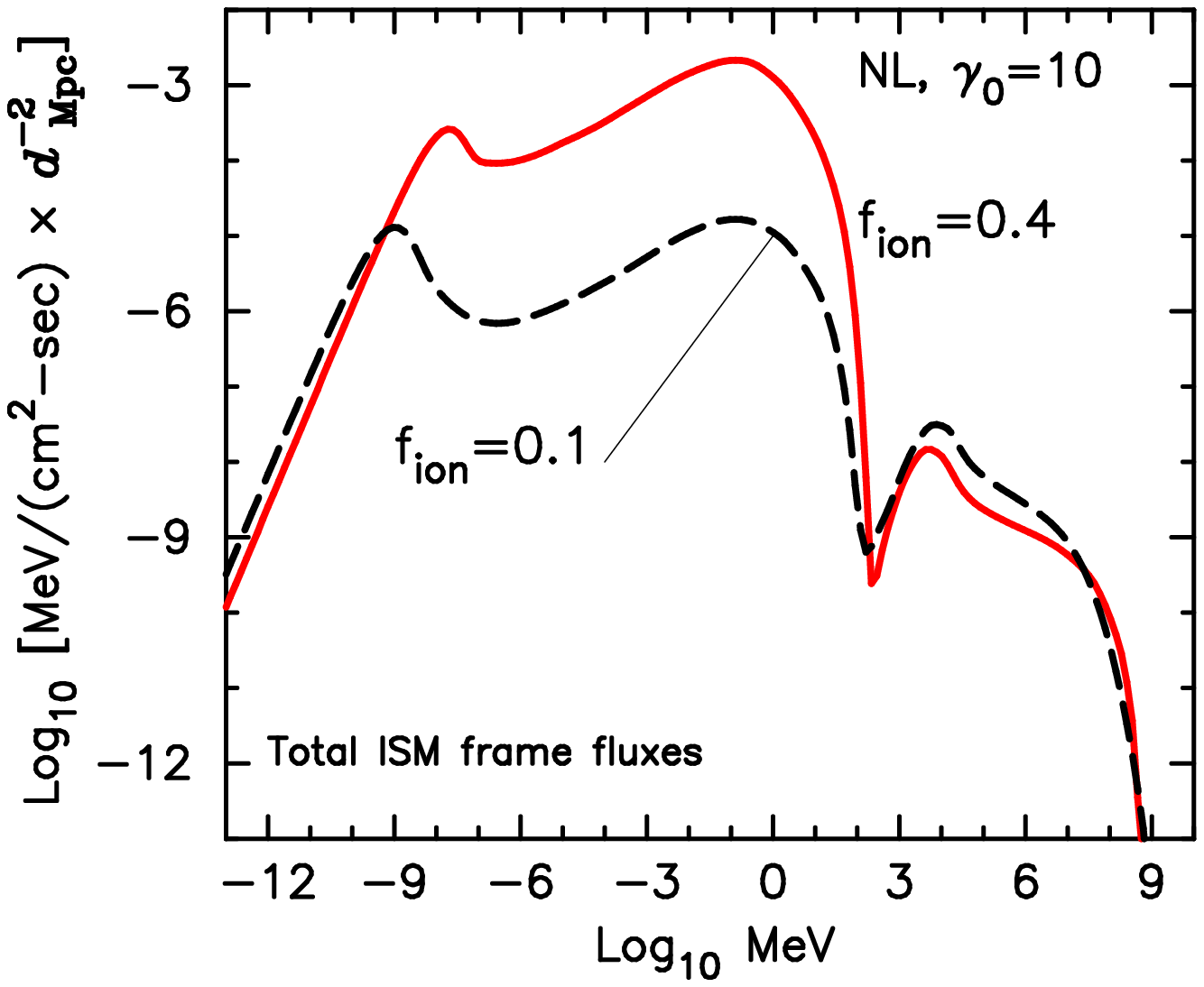}
\caption{Total observed photon energy flux for non-linear
  \mc\ models of relativistic ion-electron shocks of $\gamZ$ = 10 with
  $\Fion=0.1$ and $\Fion=0.4$ simulated by \citet{WarrenEllison2015}.
\label{fig:phot_fion}}
\end{figure}

\subsection{Radiation from \mc\ shock modeling}
Within the assumptions of the TPI model, the \mc\ technique calculates
the thermal and nonthermal plasma consistently.  Therefore, the
distribution functions, such as those shown in
Fig.~\ref{fig:fig11_mfp}, contain the background plasma as well as the
accelerated population.
This means the ion and electron spectra are absolutely normalized
for a given set of environmental and model parameters and the \syn,
\brems, and \pion\ emission is also absolutely normalized without
requiring additional parameters.
The current \mc\ code does not include \SSC\ so an external photon
field is required for \IC. Additional parameters will also be required
for the other radiation mechanisms if radiation is calculated from
shock accelerated particles interacting with an external medium.

In Fig.~\ref{fig:dNdp_fion}, distributions from
\citet{WarrenEllison2015} are shown where the fraction of energy
transferred from ions to electrons, $\Fion$, has been set to two
different values as indicated. In both cases, the NL shock structure
was calculated including electrons, protons, and He$^{2+}$.
Fig.~\ref{fig:phot_fion}, also from \citet{WarrenEllison2015}, shows
the broadband continuum radiation produced in these shocks.
The transfer of ion energy to electrons radically influences the
\syn\ emission. The emission at $\sim 1$\,Mev  with $\Fion=0.4$ is
$\sim 100$ times that for $\Fion=0.1$. On the other hand, the \pion\
emission in the GeV-TeV region drops for $\Fion=0.4$ since the ions
have less energy. See \citet{WarrenEllison2015} for a full
discussion of these effects.

%%%%%%%%%%%%%%%%%%%%%%%%%%%%%%%%%%%%%%%%%%%%%%%%%%%%%%%
\section{Summary}\label{sec:sum}

In this article, we have provided an analytical description of the
main components of a relativistic collisionless shock, building on
earlier work and borrowing from the results of particle-in-cell
simulations. We have completed this discussion with a review of
\mc\ techniques, which offer a complementary method to approach the
mesoscale physics of these shock waves.  As an executive summary, we
review here some of the most salient points of the physics of
relativistic collisionless shock waves:

\newlistroman

\listromanDE In unmagnetized shocks where the background magnetic
field is dominated by self-generated turbulence, the shock forms
thanks to a partial reflection of the incoming particles on a front
produced by the enhancement of the electromagnetic energy density.

\listromanDE The electromagnetic precursor is generated by a streaming
    micro-instability. The Weibel instability emerges as the leading
    micro-instability in the unmagnetized regime. At weak but finite
    magnetization, a current driven instability is expected to play an
    important role as well.

\listromanDE At magnetization levels not far below unity (and higher),
the shock is formed by the compression of the background magnetic
field.

\listromanDE As we have argued, the Weibel turbulence which is excited
in the precursor propagates forwards at a sub-relativistic speed
(typically $\xi_{\rm cr}$ with respect to the background plasma,
$\xi_{\rm cr}$ being the ratio of the cosmic ray pressure over the
incoming energy density in the front frame).

\listromanDE Electron heating in the precursor occurs through two
processes: through the motion of the scattering centers with respect
to the background, and through the induction electric field.

\listromanDE The motion of the Weibel frame with respect to the
  background, in addition to its heating effect, produces a slowing of
  the incoming plasma that might reduce its speed to a mildly
  relativistic one, thereby allowing a saturation of the Weibel
  instability by the dispersion effect of the beam.

\listromanDE Weibel turbulence gives rise to filaments in the
background plasma, whereas the beam of superthermal particles does not
suffer filamentation due to its high inertia.

\listromanDE The particle scattering rate decreases with the energy of
the considered particle in proportion to $\gamma^{-2}$. This result
first established with a random phase approximation has been confirmed
by a Hamiltonian chaos investigation, and has also been confirmed by
PIC simulations.

\listromanDE Because of this fast decrease of the scattering
frequency, even a tiny mean field can stop the Fermi acceleration
process; acceleration can take place in shock waves for which the
magnetization $\sigma\,\lesssim\,\xi_B^2$, i.e.,
$\sigma\,\lesssim\,10^{-4}$. In particular, if the Weibel instability
dominates turbulence production on all scales into the non-linear
regime, protons cannot be accelerated beyond $10^{16}\,$eV in the
termination shocks of gamma-ray bursts. Electrons can be efficiently
accelerated up to $10\,$TeV or so in similar conditions and give rise
to GeV synchrotron emission at early times.

\listromanDE If \rel\ shocks are as efficient in producing
superthermal particles as suggested by PIC simulations, and often
assumed in astrophysical models, non-linear effects from efficient CR
production must be considered.

\listromanDE With efficient CR production, the shock structure,
turbulence production, Fermi acceleration, and particle injection from
the thermal background are coupled. By using phenomenological particle
scattering, \mc\ methods can address this multi-scale coupling over
large dynamic ranges in an internally consistent fashion.

\listromanDE The analysis of \rel\ shocks is substantially more
difficult than \nonrel\ ones because particle distributions are not
nearly isotropic in all frames and the diffusion approximation cannot
be used directly. Nevertheless, the underlying physics of
collisionless shock formation, particle injection, wave generation,
and Fermi acceleration must be continuous from \nonrel\ speeds through
the \transrel\ regime.  \mc\ techniques can address important
non-linear aspects of this transition.

\listromanDE A major direction for future work is to include magnetic
field amplification for \rel\ and \transrel\ shocks. Additional work
can be directed toward non-linear oblique shocks and using the
\mc\ code in astrophysical applications.

\appendix

\section{Appendices of the first part of the paper}

\subsection{Determination of the Weibel frame in a pair plasma}\label{sec:appA}

We consider the transverse perturbed motion in the linear
approximation in the background plasma and in the relativistic beam of
returning particles, assuming here a pair shock.  For a species
indexed by $\alpha$, moving at velocity $\beta_\alpha$ in a frame to
be specified later, with proper enthalpy density $h_\alpha$ and number
density $n_\alpha$, the perturbed dynamical and continuity equations
can be written as
\begin{eqnarray}
\label{eq:W0}
  c^{-1} h_\alpha \gamma_\alpha \partial_t \boldsymbol{\delta u_{\alpha,\perp}} +
\boldsymbol{\nabla_{\perp}}\delta p_\alpha & = &  n_\alpha \gamma_\alpha q_\alpha \boldsymbol{\nabla_{\perp}}
\left(\delta\Phi - \beta_\alpha \delta A_x\right) \\
c^{-1} \partial_t \delta \left(\gamma_\alpha n_\alpha\right) + \gamma_\alpha n_\alpha
\boldsymbol{\nabla_{\perp}}\cdot \boldsymbol{\delta \beta_{\alpha,\perp}} & \,=\, & 0
\end{eqnarray}
It is understood that $\delta p_\alpha$, $\boldsymbol{\delta
u_\alpha}$, $\delta n_\alpha$ are first order perturbations while
$h_\alpha$, $n_\alpha$ etc. are zeroth-order uniform and constant in
time variables. The above implicitly assumes that the
electromagnetic perturbations are characterized by $\delta A^0$ and
$\delta A^x$ only, and that $\partial_x\cdot\,=\,0$ for all
quantities.

For the sake of simplicity, we assume cold species and neglect the
pressure perturbation. The more general case of warm plasmas is
discussed further below. Then for each fluid (one fluid being composed
of $+$ and $-$ species), meaning the superthermal fluid and the
background plasma fluid, we derive from the above two equations an
uncoupled equation for the evolution of its \emph{apparent} charge
perturbation:
\begin{equation}
\label{eq:W1}
\partial_t^2 \delta \rho_\alpha \,=\, \frac{\omega_{\rm
    p \alpha}^2}{4\pi} \Delta_{\perp} \left(\delta\Phi - \beta_\alpha
\delta A_x\right) \ ,
\end{equation}
where apparent charge density means $\rho_\alpha \,\equiv\, \gamma_\alpha e
(n_{\alpha +}-n_{\alpha -})$; $\omega_{\rm p,\alpha}$ represents the plasma
frequency defined in the rest frame of the considered fluid, such that
\begin{equation}
\label{eq:W2}
\omega_{\rm p,\alpha}^2 \,\equiv\, \frac{4\pi n_\alpha e^2 c^2}{h_\alpha/n_\alpha} \ .
\end{equation}
with the index $\alpha$ running over the superthermal (returning)
particles and over the background plasma.  For a plasma temperature $T
\,\ll\, mc^2$ (cold non-relativistic plasma), $h_\alpha \,\simeq\, n_a
mc^2$ and $\omega_{\rm p\alpha}^2 = 4\pi n_\alpha e^2/m$. This is the
case of the background plasma, for which $n_\alpha$ designs the
density $n_{\rm u}$ of the ambient plasma. For a plasma of
relativistic temperature $T \,>\, mc^2$, $h_\alpha \,\simeq\, 4p_a
\,\simeq\, 4/3\, \Delta\gamma_\alpha n_\alpha mc^2$ and $\omega_{\rm
  p\alpha}^2 = 3\pi n_\alpha e^2/\left(\Delta \gamma_\alpha
m\right)$. This is the case of the plasma of returning particles for
which: $n_\alpha \,\simeq\, \xi_{\rm cr} \gamma_{\rm sh} n_{\rm u}$,
while the thermal Lorentz factor is $\Delta\gamma_\alpha \,\sim\,
\gamma_{\rm sh}$ (front frame); thus its plasma frequency is such that
$\omega_{\rm pb} \,\simeq\, \xi_{\rm cr} \omega_{\rm p}$, $\omega_{\rm
  p}$ the plasma frequency of the background plasma.

We define the Weibel frame as that in which the instability becomes
purely magnetic, i.e. where the electrostatic potential and the total
electric charge density vanish. The evolution of the total charge
density is governed by
\begin{equation}
\label{eq:W3}
4\pi \partial_t^2 \delta \rho_{\rm tot} = \Delta_{\perp}\left\{\sum_\alpha\omega_{\rm p \alpha}^2\left(
\delta\Phi - \beta_\alpha  \delta A_x\right)\right\}  \ .
\end{equation}
In the Lorentz gauge, $\delta \Phi$ can be expressed directly in terms
of the total charge perturbation; switching over to Fourier space, one infers
\begin{equation}
\label{eq:W4}
\left(\frac{\omega^4}{k^2}-\omega^2+  \sum_\alpha \omega_{\rm p \alpha}^2\right)
\delta\Phi \,=\, \sum_\alpha \beta_\alpha \omega_{\rm p \alpha}^2 \delta A_x\ .
\end{equation}
Thus there is no source of electrostatic excitation when $\sum_\alpha
\beta_\alpha \omega_{p\alpha}^2 = 0$; this defines the Weibel frame, that always
exists. For the case we are treating, the relation defining the Weibel
frame is simply:
\begin{equation}
\label{eq:W5}
\xi_{\rm cr} \beta_{\rm b \vert w} + \beta_{\rm bg \vert w} = 0 \ .
\end{equation}
with $\beta_{\rm b \vert w}$ (resp. $\beta_{\rm bg \vert w}$) the
velocity of the beam (resp. the background plasma) relative to this
Weibel frame. The beam moves relativistically with respect to the
background plasma, therefore either $\beta_{\rm b \vert w}\,\sim\,1$
or $\beta_{\rm bg \vert w}\,\sim\,-1$; we will argue further on that
the plasma filamentation is carried by the background, which motivates
the former choice. Thus the Weibel frame moves forward (in the same
direction as the shock front) relative to the background plasma, but
at a low velocity
\begin{equation}
  \label{eq:Wvel}
\beta_{\rm w\mid bg} = \frac{\xi_{\rm cr}}{1+\xi_{\rm cr}} \ .
\end{equation}
One can also directly calculate the speed of the Weibel frame with
respect to the front, written $\beta_{\rm w}$ (with other velocities
are also written in the front frame):
\begin{equation}
\label{eq:Wvelf1}
\frac{\beta_{\rm w} - \beta_{\rm bg}}{1-\beta_{\rm w} \beta_{\rm bg}} = \xi_{\rm cr} \frac{\beta_{\rm b} - \beta_{\rm w}}{1- \beta_{\rm b} \beta_{\rm w}} \ ,
\end{equation}
which leads to
\begin{equation}
\label{eq:Wvelf}
\beta_{\rm w} \,\simeq\, - \beta_{\rm sh} + \frac{\xi_{\rm cr}}{\gamma_{\rm sh}^2} \ .
\end{equation}
It can be easily checked that the latter result is also obtained by a
Lorentz transform of the result in the background frame.

\subsection{Determination of the Weibel frame in a ionic plasma with hot electrons}\label{sec:appB}

We assume now a plasma made of protons and hot electrons. When
electrons are relativistically hot, they are in a quasi-Boltzmann
equilibrium and also they are responsible for strong Landau effects in
the electromagnetic response functions. The transverse and
longitudinal response functions of background electrons are defined
through
\begin{eqnarray}
\delta j_{x,e} & = & \frac{\omega^2}{4\pi c} \chi_{{\rm T},e} \delta A_{x\vert\rm u} \\
\delta \rho_e & = & -\frac{k^2}{4\pi} \chi_{{\rm L},e} \delta\Phi_{\vert\rm u}
\end{eqnarray}
with
\begin{eqnarray}
  \chi_{{\rm T},e} & \,\simeq\, & - \frac{\omega_{\rm pi}^2}{\xi_{\rm th} k^2c^2}
  \left(1-i\frac{\pi}{4}\frac{kc}{\omega}\right) \\
  \chi_{{\rm L},e} & \,\simeq\, & \frac{\omega_{\rm pi}^2}{\xi_{\rm th} k^2c^2}
  \left(1+i\frac{\pi}{2}\frac{\omega}{kc}\right)
\end{eqnarray}
and let us recall the notation: $\xi_{\rm th}\,\equiv\,kT_e/(m_p
c^2)$. Using a cold response for the background ions in their rest
frame, i.e. $\delta j_{x,i}\,=\,-\omega_{\rm pi}^2/(4\pi)\delta
A_{x\vert\rm u}$, inserting these responses in the equation for
$\delta A_{x\vert\rm u}$, one obtains the dispersion relation:
\begin{equation}
\label{eq:wg3}
\left(\omega^2 -k^2\beta_*^2c^2\right)\left(1+\frac{k^2c^2}{\omega_{\rm pi}^2} -
i\frac{\pi}{4\xi_{th}}\frac{\omega}{kc}\right)\,=\,
-\xi_{\rm cr} \beta_{\rm b\vert u}^2 k^2c^2 \ .
\end{equation}
This dispersion relation neglects terms of order $\omega^2/k^2$ in the
second term on the l.h.s., and it neglects a contribution of the
electrostatic potential on the r.h.s, which is of order $\beta_{\rm
  w\vert u}\,\sim\,\xi_{\rm cr}$ relative to the above term.  The
effective sound velocity $\beta_*c$ which enters the equation of
motion for the charge perturbation of the beam is of order
$1/\gamma_{\rm b}$, i.e.  $\beta_*^2c^2 = c_{\rm s,b}^2/\gamma_{\rm
  sh}^2$ with $c_{\rm s,b}=1/\sqrt{3}$.

Let us write the complex frequency of the Weibel modes as follows:
$\omega = iz \xi_{\rm th} kc$, where $z \in \mathbb{C}$. The
modification of the transverse response by the electronic Landau
effect is the main modification of the dispersion relation for Weibel
modes, at least off the saturation regime, that reads (omitting terms
of order $1/\gamma_{\rm sh}^2 \xi_{\rm cr}^2$):
\begin{equation}
\label{eq:wg4}
\xi_{\rm th}^2\frac{\pi}{4} z^3 + a\xi_{\rm th}^2 z^2 + \frac{\pi}{4}
\beta_*^2 z + a\beta_*^2 - \xi_{\rm cr} = 0 \ ,
\end{equation}
where $a \equiv 1 +k^2c^2/\omega_{\rm pi}^2$. The physical situation of interest
corresponds to $\xi_{\rm cr} \gg  \xi_{\rm th}$ where
\begin{equation}
\label{eq:wg5}
z^3 \,\simeq\, \frac{4}{\pi} \frac{\xi_{\rm cr}}{\xi_{\rm th}^2} \ .
\end{equation}
We thus find an instability with a growth rate:
\begin{equation}
\label{eq:wg6}
g_{\rm inst\vert u} \,=\, \left(\frac{4 \xi_{\rm cr}}{\pi \xi_{\rm th}^2} \right)^{1/3} \xi_{\rm th} kc
\,\simeq\, \left(\xi_{\rm th} \xi_{\rm cr}\right)^{1/3} kc \ .
\end{equation}
This growth rate is not significantly different from the cold one for
$kc = \omega_{\rm pi}$ and $\xi_{\rm th} \,\sim\, \xi_{\rm cr}$. The
threshold due to beam dispersion is found by keeping all the terms
except the third degree term and assuming $\xi_{\rm cr} - a\beta_*^2$
positive but arbitrary small. Then we find the unstable root:
\begin{equation}
\label{eq:wg7}
z \,\simeq\, \frac{4}{\pi \beta_*^2} \left(\xi_{\rm cr} - a \beta_*^2\right) \ ,
\end{equation}
which leads to the saturation condition: $\gamma_{\rm bg} = \xi_{\rm
  cr}^{-1/2}$ in the front frame.

\subsection{The scattering law derived from the Hamiltonian chaos}\label{sec:appC}

We thus consider returning cosmic rays in the Weibel frame with a very
high Lorentz factor $\gamma$, such that their motion in the
$x-$direction is almost constant ($v_x =c$ with a very good accuracy);
then, only their non-relativistic transverse motion matters with
$v_{\perp} \,\sim\, c/\gamma_{\rm sh}$. We assume that the magnetic
modes carry an $x-$wave number $k_{0}$ which is small compared to the
transverse ones $k_{0} \,\ll\, k_{\perp} \,\sim\, \delta_i^{-1}$, as
it appears to be in numerical simulations. Thus the cosmic rays can
suffer resonances with Weibel modes: $k_{\perp} v_{\perp} \pm k_0c =
0$.

The most simple description consists in considering the transverse
dynamics only, the contribution of the x-coordinate amounting to a
time dependent driving with $k_0x(t) = k_0c t$, with a frequency
$\omega_0 \,\equiv\, k_0c \,\ll\, \omega_{pi}$. In a 2D description,
with $y$ denoting the transverse coordinate, this simple description
involves a Lagrangian of the form:
\begin{eqnarray}
\label{eq:lag}
{\cal L} & \,=\,& \frac{\dot y^2}{2} + \varepsilon \cos \left(\omega_0t\right) \sum_n
a_n \cos \left(k_ny\right) \nonumber\\
& \,=\, & \frac{\dot y^2}{2} + \frac{\varepsilon}{2}
\sum_n a_n\left[\cos \left(k_ny-\omega_0t\right) + \cos\left(k_ny + \omega_0t\right)\right] \ ,
\end{eqnarray}
where the small amplitude parameter is defined as
\begin{equation}
  \varepsilon \,\equiv\, \frac{q\delta A_{x\vert\rm w}}{\gamma mc^2}
\end{equation}
The corresponding Hamiltonian describes a Chirikov standard map, with
the small $\epsilon$ parameter characterizing the perturbation which
will provide the transition to chaotic dynamics.

Let us define a local coordinate of phase space $\theta_n \,\equiv\,
k_ny(t) \mp \omega_0t$. A resonance is defined by $\dot \theta_n =
0$. Around such a resonance, the Lagrangian can be approximated as the
time average:
\begin{equation}
\label{eq:lag2}
\bar {\cal L}_n \,=\, \frac{1}{2k_n^2} \left(\dot \theta_n \pm \omega_0\right)^2 +
     \frac{\varepsilon}{2} a_n \cos \theta_n \ .
\end{equation}
The conjugate momentum of $\theta_n$ is
\begin{equation}
\label{eq:lag3}
J_n \,\equiv\, \frac{\partial \bar {\cal L}_n}{\partial \dot \theta_n} \,=\,
\frac{\dot \theta_n \pm \omega_0}{k_n^2} \ .
\end{equation}
and the corresponding Hamiltonian is:
\begin{equation}
\label{eq:lag4}
\bar H_n \,=\, \frac{k_n^2}{2} J_n^2 \mp \omega_0 J_n - \frac{\varepsilon}{2} a_n \cos \theta_n \ .
\end{equation}
This time independent Hamiltonian describes the energy invariant
around each resonance. The $n-$resonance occurs for $J_n \,=\,
J_n^{(r)} \,=\, \pm \omega_0/k_n^2$ (which is nothing but $k_n v_n =
\pm \omega_0$). Assuming for the time being that the resonances are
sufficiently isolated, then setting $J_n\,=\,J_n^{(r)} +\delta J_n$
around such a resonance, one finds a pendulum like Hamiltonian, with
oscillations described by
\begin{equation}
\label{eq:lag5}
\delta \dot J_n \,=\, \frac{\ddot \theta_n}{k_n^2} = - \frac{\varepsilon}{2} a_n \sin \theta_n \ .
\end{equation}
The oscillations in $\theta_n$ have a pulsation $\Omega_n$ in the
bottom of the well such that:
\begin{equation}
\label{eq:lag6}
\Omega_n^2 \,\equiv\, \frac{\varepsilon}{2} k_n^2 a_n \ .
\end{equation}
The separatrix being defined as the region which separates
oscillations from rotations, one derives the maximum amplitude of
the separatrix around the $n-$resonance in the $J_n$ direction of
phase space as $\Delta J_n = 2\sqrt{2}\Omega_n/k_n^2$. In contrast,
the distance between two neighboring resonances is
$\omega_0\left(k_n^{-2} - k_{n-1}^{-2}\right) \,\simeq\, 2\omega_0
\delta k_n/k_n^3$. The simple heuristic threshold for chaos,
proposed by \citet{1979PhR....52..263C}, is that the separatrix
amplitude be larger than the spacing between resonances, giving:
$\Omega_n > \omega_0 \delta k_n / k_n^2$, which is always easily
satisfied in our case, the transverse spectrum of Weibel instability
being a continuum.

Therefore, above a very low threshold, the resonances overlap and the
transverse motions of cosmic rays randomly jump from one resonance to
another one. Using Eq.~(\ref{eq:lag5}), one infers the evolution of
$J_n$ after $i$ jumps:
\begin{equation}
\label{eq:lag7}
\left\langle\Delta J_n^2\right\rangle \,\sim\, \frac{\varepsilon^2}{8}a_n^2\Delta t^2 i  \ ,
\end{equation}
with $\Delta t \,\simeq\, \omega_0^{-1}$ the time taken to change
resonance, which plays the role of coherence time. The above assumes a
random $\theta_n$ in resonance $n$ at step $i$. Now, the variation in
$J_n$ is related to the variation of the pitch angle $\alpha$ of the
particle, since $\Delta J_n\,=\,\Delta v_\perp/k_n\,\simeq\, \Delta
\alpha/k_n$. As $t\,=\,i\,\Delta t\,\rightarrow +\infty$, one thus
infers a scattering frequency
\begin{equation}
  \nu_s \,=\,\frac{\langle\Delta\alpha^2\rangle}{2t}\,\propto\,
  \varepsilon^2 k_n^2 \,\propto\, \xi_B \omega_{\rm pi}
  \frac{\gamma_{\rm sh}}{\gamma^2}\ .
\end{equation}

\begin{acknowledgements}
G. Pelletier and M. Lemoine were financially supported by the
Programme National Hautes \'Energies (PNHE) of the C.N.R.S. and by the
ANR-14-CE33-0019 MACH project. A. M. Bykov was supported by RSF grant
16-12-10225.
\end{acknowledgements}

\bibliographystyle{aps-nameyear}
\bibliography{bib_comb4}
\end{document}